\documentclass[pre,twocolumn]{revtex4}
\usepackage{amsmath,amsfonts,epsfig}
\usepackage[english]{babel}
\usepackage{graphicx}

\newcommand{\figwidth}{\columnwidth}
\newcommand{\miniwidth}{\columnwidth}

\begin{document}

\title{Effective shape and phase behaviour of short charged rods}
\author{Eelco Eggen$^1$, Marjolein Dijkstra$^2$, and Ren\'{e} van Roij$^1$}
\affiliation{$^1$Institute for Theoretical Physics, Utrecht
University, Leuvenlaan 4, 3584 CE Utrecht, The
Netherlands\\$^2$Soft Condensed Matter, Debye Institute for
Nanomaterials Science, Utrecht University, Princetonplein 5, 3584
CC Utrecht, The Netherlands}
\date{8 July 2008}

\begin{abstract}
We explicitly calculate the orientation-dependent second virial
coefficient of short charged rods in an electrolytic solvent,
assuming the rod-rod interactions to be a pairwise sum of
hard-core and segmental screened-Coulomb repulsions. From the
parallel and isotropically averaged second virial coefficient, we
calculate the effective length and diameter of the rods, for
charges and screening lengths that vary over several orders of
magnitude. Using these effective dimensions, we determine the
phase diagram, where we distinguish a low-charge and
strong-screening regime with a liquid crystalline nematic and
smectic phase, and a high-charge and weak-screening regime with a
plastic crystal phase in the phase diagram.
\end{abstract}

\maketitle

\section{Introduction}
The study of suspensions of non-spherical colloidal particles
started with the experimental works of \citet{Zocher} and
\citet{Bawden}, and with Onsager's theoretical work
\citep{Onsager}. It has since developed into a very versatile
field of research. A lot of attention has been focussed on
needle-shaped rods, either naturally occurring ones such as
viruses like Tobacco Mosaic Virus or fd virus
\cite{Bawden,Fraden,Dogic}, or laboratory-synthesized ones such as
Boehmite rods \citep{Bruggen}. In recent years, however, a
plethora of non-spherical particles have been synthesized that are
not extremely elongated, for example: ellipsoidal colloids with
aspect ratio $\sim 3$ \citep{Snoeks}; colloidal dumbbells
\citep{Johnson}; or nano-particles with the shape of a rod, disk,
snowman, cube, cap, or raspberry,
\citep{Yin,Sheu,Jun,Murphy,Zoldesi,Cho,Kraft}. These particles are
often charged when dissolved in a polar solvent like water, and
hence their pair-interactions involve not only the anisotropic
steric short-range repulsions but also electrostatic long-range
repulsions. The strength of the latter is determined by the charge
on the particle and the range is determined by the Debye screening
length of the solvent \citep{Derjaguin:fara,Verwey}. For small
charges and strong screening (i.e.~high salt concentrations), one
expects the steric interactions to be dominant (if we assume that
dispersion forces can be neglected). Hence, one can use computer
simulations or theoretical studies of {\em hard} anisotropic
bodies
\citep{Bolhuis,Schilling,Cuesta,Martinez-Raton,Eppenga,Schmidt} to
get an idea of the phase diagram of the system as a function of
concentration. In the case of a high charge or weak screening
(i.e.~low salt concentration), however, the situation is less
clear-cut. There, the degree of anisotropy of the electrostatic
interactions is not obvious from the outset: on the one hand one
expects the soft screened-Coulomb interactions to wash out the
hard-core anisotropy such that the interactions become effectively
more spherically symmetric, while on the other hand there are the
intriguing findings reported for example in
Refs.~\citep{Ramirez,Trizac}. The studies in these papers apply to
systems of charged anisotropic particles in a screening medium. It
was found that the electrostatic anisotropy persists to infinitely
large distances as the asymptotic decay of each multipole
contribution to the electrostatic potential due to a nonspherical
charge distribution is equal \citep{Ramirez,Trizac}. This
conclusion is in sharp contrast to the case of a charge
distribution in vacuum, where the monopole potential decays more
slowly than that of the dipole, as each order multipole
contribution decays slower than the next one does. In our paper we
investigate the interplay between hard-core and electrostatic
interactions for non-spherical particles, for the relatively
simple particle shape of spherocylinders.

It is well-established by now that non-spherical colloidal
particles can form a wealth of phases in thermodynamic
equilibrium. Needle-like colloidal rods, for instance, form a
phase sequence I--N--Sm--X upon increasing the concentration from
very dilute up to close packing, where I is the completely
disordered isotropic fluid phase, N the liquid crystalline nematic
phase with orientational ordering, Sm the smectic-A phase built
from orientationally ordered liquid-like layers, and X a fully
ordered crystal phase
\citep{Zocher,Onsager,Veerman,Vroege,Bolhuis,Fraden,Poniewierski,Tarazona,Graf:DFT}.
This phase sequence for colloidal needles is well-established for
hard-core interactions \citep{Bolhuis,Poniewierski,Tarazona}.
Also, for softer electrostatic screened-Coulomb repulsions in the
case of charged needles, at least in the regime where the length
of the rods is much larger than the diameter and the screening
length of the electrolytic solvent. This ensures that the
effective diameter of the rods is much smaller than the length
\citep{Stroobants,Fraden}. By contrast, particles with shapes that
are sufficiently close to spherical are {\em not} expected to
exhibit the liquid crystalline phases N and Sm due to their small
anisotropy. Instead, for such near-spherical particles one would
expect a plastic crystalline phase (P) to appear in the phase
diagram, residing in between the isotropic fluid and the fully
ordered crystal. The P phase is characterized by positional
ordering on a lattice, but without long-ranged orientational
ordering of the particles. For instance, a phase sequence I--P--X
upon increasing the concentration has indeed been established in
simulations of short hard spherocylinders and of hard dumbbells
with a length-to-diameter ratio smaller than about 0.35
\citep{Bolhuis,Vega,Marechal}. The question we address in this
paper concerns the effect of colloidal charge and ionic screening
on the effective shape of relatively short rods, and on their
expected phase sequence upon increasing the concentration. On the
basis of the well-established increase of the effective diameter
of charged needles compared to their hard-core diameter
\citep{Stroobants}, it is to be expected that high colloidal
charges and weak-screening conditions (i.e.~low salt
concentrations) lead to a decreased anisotropy of short charged
rods. Hence, this will lead to a larger tendency of the system to
exhibit a plastic crystal phase instead of liquid crystalline
phases in the phase diagram, even if the hard-core shape would
allow for liquid crystalline equilibrium phases.

Of course, suspensions of charged rods have been extensively
studied theoretically before. Many of these studies are based on
Onsager's second virial theory for hard rods \citep{Onsager},
which is modified and extended to take into account the effects of
charge and screening on the isotropic-to-nematic transition
\citep{Stroobants,Sato,Nyrkova:bio,Nyrkova:mac,Chen-Koch,Potemkin}.
Some of these studies, for example those of
Refs.~\citep{Onsager,Stroobants,Sato}, focus on the needle limit
in which the rod length is very large compared to the screening
length. In this limit, only the diameter is affected by the
electrostatic effects, but in such a way that the effective
geometry of the rod remains needle-like. In
Refs.~\citep{Nyrkova:bio,Nyrkova:mac} rod lengths of the order of
(or larger than) the Debye length are considered, at the expense,
however, of ignoring many of the prefactors such that the theory
is essentially a scaling theory. Interestingly, this scaling
theory predicts nematic--nematic coexistence in some parameter
regime, which was later confirmed in Ref.~\citep{Chen-Koch}. This
coexistence regime is characterized by a small rod charge density,
such that the effective geometry of the rod is no longer
needle-like. Another limit that was studied in detail is the limit
of weak electrostatic interactions, which naturally leads to a
perturbative description \citep{Klein,Chen-Koch,Graf:TMV}. These
schemes are very successful at describing the effective
(non-needle-like) geometry that shows up in the angular dependence
of the second virial coefficient. Another very interesting effect
was identified in Ref.~\citep{Potemkin}, where the correlation
free energy of the many-body system of charged rods and
counterions was calculated, resulting in an enhanced tendency to
orientational ordering and also the possibility of
nematic--nematic coexistence. With the notable exception of
Ref.~\citep{Graf:TMV}, however, most of these works on charged
rods focus on the isotropic and nematic phases and hence,
implicitly, on rods which are sufficiently elongated to give
liquid crystalline phases at all.

In this paper we take a slightly different perspective. We
explicitly calculate the orientation-dependent second virial
coefficient of rather short charged rods numerically, for
colloidal charges and screening lengths that vary over many
decades. Such calculations, in which we use expansions in
spherical harmonics, do not require only the asymptotic far-field
expressions of the multipoles (such as considered in
Refs.~\citep{Ramirez,Trizac}), but in fact their full distance
dependence. From the resulting second virial coefficient, we
determine an effective hard-core length and diameter.
Subsequently, we use these|in combination with the published
hard-core phase diagram \citep{Bolhuis}|to determine the expected
phase sequence upon increasing the concentration. This scheme is
too crude to distinguish subtleties such as whether or not there
is a nematic-nematic coexistence regime or to what extent the
isotropic-nematic phase gap is affected. However, it is supposed
to indicate reliably whether liquid crystalline (N and Sm) or
plastic crystal (P) phases are to be expected in between the
isotropic (I) and crystalline (X) phase. We focus on the case
where the rod length is of the order of the screening length or
smaller, in contrast to most of the previous theoretical work.
This is the regime where the crossover from N and Sm to P is
expected to occur. In the limit where the rod length is small and
the hard-core interactions are important, we give a simplified
theoretical description that turns out to be in remarkable
agreement with the numerical results. As our numerical approach
relies on an expansion in spherical harmonics of the effective
pair interaction between two rods, it leads to explicit but
involved expressions. We present some of the mathematical
technicalities of the derivation of these expressions in the
appendix.

\section{Model}
We consider a system of identical charged colloidal rods suspended
in an electrolyte solvent of dielectric constant $\epsilon$, Debye
screening length $\kappa^{-1}$, and Bjerrum length $l_{\rm
B}=e^{2}/(4\pi\epsilon k_{\rm B}T)$, at temperature $T$. Here $e$
is the elementary charge, and $k_{\rm B}$ is the Boltzmann
constant. The rods are assumed to have the shape of a
spherocylinder consisting of a cylinder of length $L$ and diameter
$D$ capped by two hemispheres also of diameter $D$. The rods have
a fixed charge, which we treat here as an (effective) line-charge
density $e\lambda$ distributed homogeneously on the axis of the
cylinder. We are interested in the effective pair potential
$V(\mathbf{r};\hat{\omega},\hat{\omega}')$ between two rods with
orientations $\hat{\omega}$ and $\hat{\omega}'$ at a
center-to-center vector $\mathbf{r}$, thermally averaged over the
degrees of freedom of the electrolyte solvent (characterized by
$\kappa^{-1}$ and $l_{\rm B}$).  In the spirit of Derjaguin,
Landau, Verwey and Overbeek (DLVO), we assume that the effective
pair potential consists of steric hard-core repulsions and
electrostatic screened-Coulomb interactions between segments of
the line charge of the two rods. We ignore short-ranged Van der
Waals attractions (i.e.~we assume the particle and the solvent to
be index-matched or that the dispersion forces are cancelled by
steric or charge stabilization). Within these approximations the
effective pair potential can be written as
\begin{equation}\label{eq:pair}
\beta V(\mathbf{r};\hat{\omega},\hat{\omega}') = \left\{
\begin{array}{l}
\infty \qquad \mbox{for overlapping rods,}\\
\\
\beta V_{\rm e}(\mathbf{r};\hat{\omega},\hat{\omega}') \hfill
\mbox{otherwise,}
\end{array}\right.
\end{equation}
where $\beta^{-1}=k_{\rm B}T$, the overlap refers to the hard-core
repulsions, and the electrostatic interaction potential is given
by
\begin{align}
\beta V_{\rm e}&(\mathbf{r};\hat{\omega},\hat{\omega}')\nonumber\\
&{}= l_{\rm B}\lambda^{2} \int_{-\frac{L}{2}}^{+\frac{L}{2}}{\rm
d}l \int_{-\frac{L}{2}}^{+\frac{L}{2}}{\rm d}l'\,
\frac{\exp[-\kappa|\mathbf{r}+l'\hat{\omega}'-l\hat{\omega}|]}{|\mathbf{r}+l'\hat{\omega}'-l\hat{\omega}|}.
\label{eq:electro}
\end{align}
The integration variables $l$ and $l'$ play the role of
coordinates running along the cylinder axis of each of the two
rods, from one end of the cylinder to the other end. In the
long-rod limit, $L/D\gg 1$ and $\kappa L\gg 1$, one can replace
the integration domains in equation \eqref{eq:electro} by the full
real axis, together with the constraint that the cylinder axes are
in ``cross configuration'' (i.e.~the axes intersect when projected
onto the plane parallel to both axes). Otherwise, the potential
vanishes. One then easily shows that $V_{\rm e}$ only depends on
the shortest distance and the relative angle between the two rods
\citep{Onsager,Stroobants,Chen-Koch}. Here we focus on shorter
rods, for which this simplification does not apply. In the
appendices we derive systematic series expansions in spherical
harmonics to describe the angular and position dependence of
$V_{\rm e}$ explicitly, focussing on rods that are rather short
compared to the Debye screening length (which sets the range of
the electrostatic repulsions). More specifically, the expansion of
the angular dependence is truncated and we consider each term as
an expansion in $\kappa L$ up to fourth order (see appendix). We
compare the result with the large-$\kappa L$ limit.

The present model can be characterized by a few dimensionless
combinations. In the limit of uncharged rods ($\lambda=0$), the
aspect ratio $L/D$ of the hard-core dimensions is of primary
importance. However, for the charged rods of present interest, the
ratio $\kappa L$ (of the hard-core length to the Debye screening
length of the solvent) gives more information on the interaction
anisotropy. The ratio $\kappa D$ is relevant as a measure of ionic
strength. Dimensional inspection of the expression in equation
\eqref{eq:electro} shows that the strength of the electrostatic
interactions is determined by the dimensionless (square of the)
line charge density
\begin{equation}
q \equiv \frac{l_{\rm B}\lambda^{2}}{\kappa}.
\end{equation}
These dimensionless combinations can span quite a range of
numerical values in experimental systems. For instance, for fd
virus suspended in water one finds \citep{Dogic} $L/D>100$,
$\kappa D\simeq 0.1$--$1$, and $q=70$--$700$, and recently
synthesized silica dumbbells in oily solvents \citep{Demirors} are
best characterized by $L/D\simeq 1$, $\kappa D\simeq 1$, and
$q\simeq 10^2$. Short (double stranded) DNA chains have $\kappa
D\simeq 0.1$--$1$ and $q\simeq 0.1$--$10$, while their length can
be varied by the number of base pairs included in the sequence.
These chains can be characterized as rigid rods up to the
persistence length corresponding to $L/D\simeq 50$. Moreover,
present-day synthesis techniques allow for the tuning of surface
charge, in principle at least, from essentially vanishing to
extremely high. This is achieved for example by using different
coatings with varying degrees of ion-dissociation of the surface
groups. It is therefore of interest to investigate the
thermodynamics of the present model over a wide range of
parameters.

\section{Thermodynamics and Effective dimensions}
With the pair potential specified by equations \eqref{eq:pair} and
\eqref{eq:electro}, and with an explicit scheme to evaluate it as
explained in the appendix, we can study the macroscopic properties
of suspensions of these charged rods. In principle, we do this as
a function of concentration, for various $q$, $\kappa D$, and
$L/D$. Here we circumvent the complexity of the full
statistical-mechanical calculation of free energies and phase
diagrams of the system at hand. We do this by mapping the second
virial coefficient of the {\em charged} spherocylinders of
interest onto that of {\em hard} spherocylinders with an effective
cylinder length $L_{\rm eff}$ and an effective diameter ${D_{\rm
eff}}$ that we will calculate below. We then {\em presume} that
the phase diagram of the system of charged rods follows from that
of the effective hard-rod system, which we take from published
computer simulation data \citep{Bolhuis}. It is well-known from
these and follow-up simulations of hard rods, as well as density
functional theory \citep{Poniewierski,Tarazona}, that this system
exhibits a sequence of phase transitions upon increasing the
concentration that strongly depends on the aspect ratio $L/D$:
sufficiently elongated hard rods with $L/D>3.7$ have a phase
sequence isotropic--nematic--smectic--crystal (I--N--Sm--X),
sufficiently short hard rods $0<L/D<0.35$ show a sequence I--P--X
with P a plastic crystal, and in between there are two more
regimes in which the N and P phase, respectively, do no longer
appear in the phase sequence. Below we determine how the analogous
crossovers between these regimes of the effective system, as
determined by $L_{\rm eff}/D_{\rm eff}$, depend on the parameters
$q$, $\kappa D$, and $L/D$.

\begin{figure}[t]
\includegraphics[width=\figwidth]{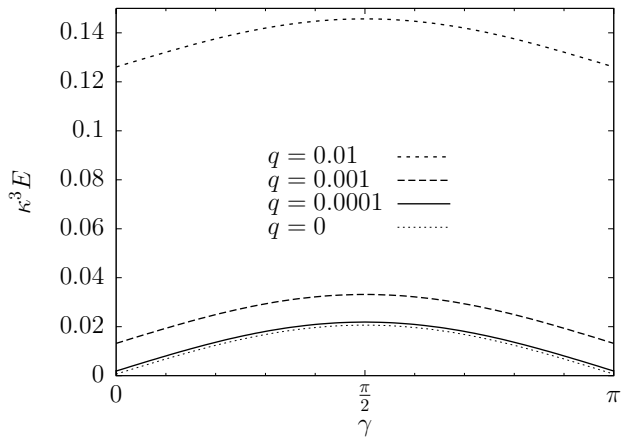} \caption{The effective
excluded volume $\kappa^{3}E$, as a function of the angle $\gamma$
between the two rod orientations, for different values of the
charge parameter $q$. We used the parameter values $\kappa L=1$
and $\kappa D=0.01$ (such that $L/D=100$).} \label{fig:excl-vol}
\end{figure}

A key ingredient of our calculation is the effective excluded
volume $E(\hat{\omega},\hat{\omega}')$ of two charged rods with
orientations $\hat{\omega}$ and $\hat{\omega}'$, defined as
\begin{equation}\label{eq:E}
E(\hat{\omega},\hat{\omega}') = \int{\rm
d}\mathbf{r}\,\Bigl(1-\exp\bigl[-\beta
V(\mathbf{r};\hat{\omega},\hat{\omega}')\bigr]\Bigr),
\end{equation}
where the pair potential between the rods is given in
Eqs.~\eqref{eq:pair} and \eqref{eq:electro}. Note that
$E(\hat{\omega},\hat{\omega}')$ is in fact twice the corresponding
second virial coefficient, and that the nomenclature ``effective
excluded volume'' stems from the fact that it reduces to the
actual excluded volume of the pair in the case of purely hard-core
interactions. On the basis of symmetry arguments one easily checks
that the angular dependence of $E(\hat{\omega},\hat{\omega}')$ is
in fact only through the angle
$\gamma=\arccos(\hat{\omega}\cdot\hat{\omega}')$ between the
cylinder axes of the two rods. In Fig.~\ref{fig:excl-vol} we show
this $\gamma$-dependence of $E$ for rods characterized by $\kappa
L=1$ and $\kappa D=0.01$ (so $L/D=100$ and weak screening), for
several charge parameters $q$ ranging from $q=0$ (uncharged) to
$q=0.01$ (fairly charged). The results of Fig.~\ref{fig:excl-vol}
stem from a combination of numerical and analytic procedures
explained in detail in the appendix. These involve a five-fold
integration: over the contour of the rods $l$ and $l'$ in
Eq.~\eqref{eq:electro}, and the center-to-center separation vector
$\mathbf{r}$ in Eq.~\eqref{eq:E}.

The key observations of Fig.~\ref{fig:excl-vol}, which is typical
for many system parameters, are that for increasing $q$ the
effective excluded volume becomes (i) less anisotropic, and (ii)
larger in magnitude. Moreover, for all $q$ the effective excluded
volume is larger for perpendicular orientations than for parallel
ones. Qualitatively, and in fact quantitatively for many
parameters, this behaviour is identical to that of {\em hard}
spherocylinders of effective length $L_{\rm eff}$ and diameter
$D_{\rm eff}$, for which the excluded volume is given by
\citep{Onsager}
\begin{equation}\label{Veff}
\mathcal{V}_{\rm eff}(\hat{\omega},\hat{\omega}') =
\frac{4\pi}{3}D_{\rm eff}^{3} + 2\pi L_{\rm eff}D_{\rm eff}^{2} +
2L_{\rm eff}^{2}D_{\rm eff}\sin\gamma.
\end{equation}
In principle one can fit the functional form of Eq.~\eqref{Veff}
to the numerical results such as those of Fig.~\ref{fig:excl-vol}
to determine the effective hard-core dimensions $L_{\rm eff}$ and
$D_{\rm eff}$ for given charged-rod parameters. However, instead
of fitting the full angular dependence numerically, it is more
convenient to match the isotropically-averaged effective excluded
volume and the parallel one, given by
\begin{align} E_{\rm iso} ={}&
\frac{1}{(4\pi)^{2}}\int{\rm d}\hat{\omega}\int{\rm
d}\hat{\omega}'\,E(\hat{\omega},\hat{\omega}')\nonumber\\
={}& \frac{1}{2}\int_{0}^{\pi}{\rm
d}\gamma\,\sin\gamma\,E(\gamma),\\
E_{\parallel} ={}& E(\hat{\omega},\hat{\omega})=E(\gamma=0),
\end{align}
to the values for spherocylinders with effective hard-core
dimensions $L_{\rm eff}$ and $D_{\rm eff}$
\begin{align}
\mathcal{V}_{\rm iso} ={}& \frac{4\pi}{3}D_{\rm eff}^{3} + 2\pi
L_{\rm eff}D_{\rm eff}^{2} +
\frac{\pi}{2}L_{\rm eff}^{2}D_{\rm eff},\\
\mathcal{V}_{\parallel} ={}& \frac{4\pi}{3}D_{\rm eff}^{3} + 2\pi
L_{\rm eff}D_{\rm eff}^{2},
\end{align}
respectively. This procedure yields the effective hard-core
dimensions
\begin{align}
D_{\rm eff} ={}& \left[\frac{3E_{\parallel}}{4\pi}\left(1 +
3\Delta -
\sqrt{3\Delta(2+3\Delta)}\right)\right]^{\frac{1}{3}},\label{eq:deff}\\
\frac{L_{\rm eff}}{D_{\rm eff}} ={}& 2\Delta +
\frac{2}{3}\sqrt{3\Delta(2+3\Delta)},\label{eq:ldeff}
\end{align}
where we used, for notational convenience, the dimensionless
anisotropy parameter $\Delta$ defined as
\begin{equation}\label{eq:delta}
\Delta \equiv \frac{E_{\rm iso}-E_{\parallel}}{E_{\parallel}}.
\end{equation}
It turns out that inserting $L_{\rm eff}$ and $D_{\rm eff}$ as
obtained from Eqs.~\eqref{eq:deff}, \eqref{eq:ldeff}, and
\eqref{eq:delta} into Eq.~\eqref{Veff} gives an angular dependence
that is in very good agreement with the numerically obtained
effective excluded volume of charged rods.

It is also interesting to compare our numerical results with
analytic expressions that are valid in the limit where $L/D\gg 1$
and $\kappa L\gg 1$, as obtained by \citet{Stroobants}. In this
needle-limit the effective excluded volume is given by
\begin{align}
E_{\infty}(\gamma) = 2L^{2}\kappa^{-1}\sin\gamma
&\biggl[\gamma_{\rm E} + \ln 2\pi q - \ln\sin\gamma
\nonumber\\
&{} + \Gamma\left(0,\frac{2\pi q\exp[-\kappa
D]}{\sin\gamma}\right)\biggr],
\end{align}
where $\gamma_{\rm E}\approx 0.577$ is the Euler-Mascheroni
constant and where the incomplete gamma function (or exponential
integral) is defined by
\begin{equation}
\Gamma(\alpha,x) = \int_{x}^{\infty}{\rm
d}y\,y^{\alpha-1}\exp[-y].
\end{equation}
From this expression|using the Onsager limit $\mathcal{V}_{{\rm
iso},\infty}=(\pi/2) L^{2}D_{{\rm eff},\infty}$ for the
isotropically averaged excluded volume|the effective diameter can
be calculated
\begin{align}\label{eq:stroob}
\kappa D_{{\rm eff},\infty} &{}= \gamma_{\rm E} + \ln 2\pi q + \ln
2 - \frac{1}{2}\nonumber\\
&{} + \frac{2}{\pi}\int_{0}^{\pi}{\rm
d}\gamma\,\sin^{2}\gamma\,\Gamma\left(0,\frac{2\pi q\exp[-\kappa
D]}{\sin\gamma}\right).
\end{align}
The effective length is taken equal to the rod length $L_{{\rm
eff},\infty}=L$.

\section{Numerical Results}
Calculations such as those of Fig.~\ref{fig:excl-vol} are
reasonably accurate for values of $\kappa L$ roughly up to 2. For
higher values the applied approximations become poor, such that
for $\kappa L\gtrsim 3$ the calculations become even qualitatively
unreliable for many of our parameters. For this reason we restrict
most of our attention to the regime where $\kappa L\leq 2$.

\begin{figure*}[t]
\begin{minipage}{\miniwidth}
(a)
\includegraphics[width=\columnwidth]{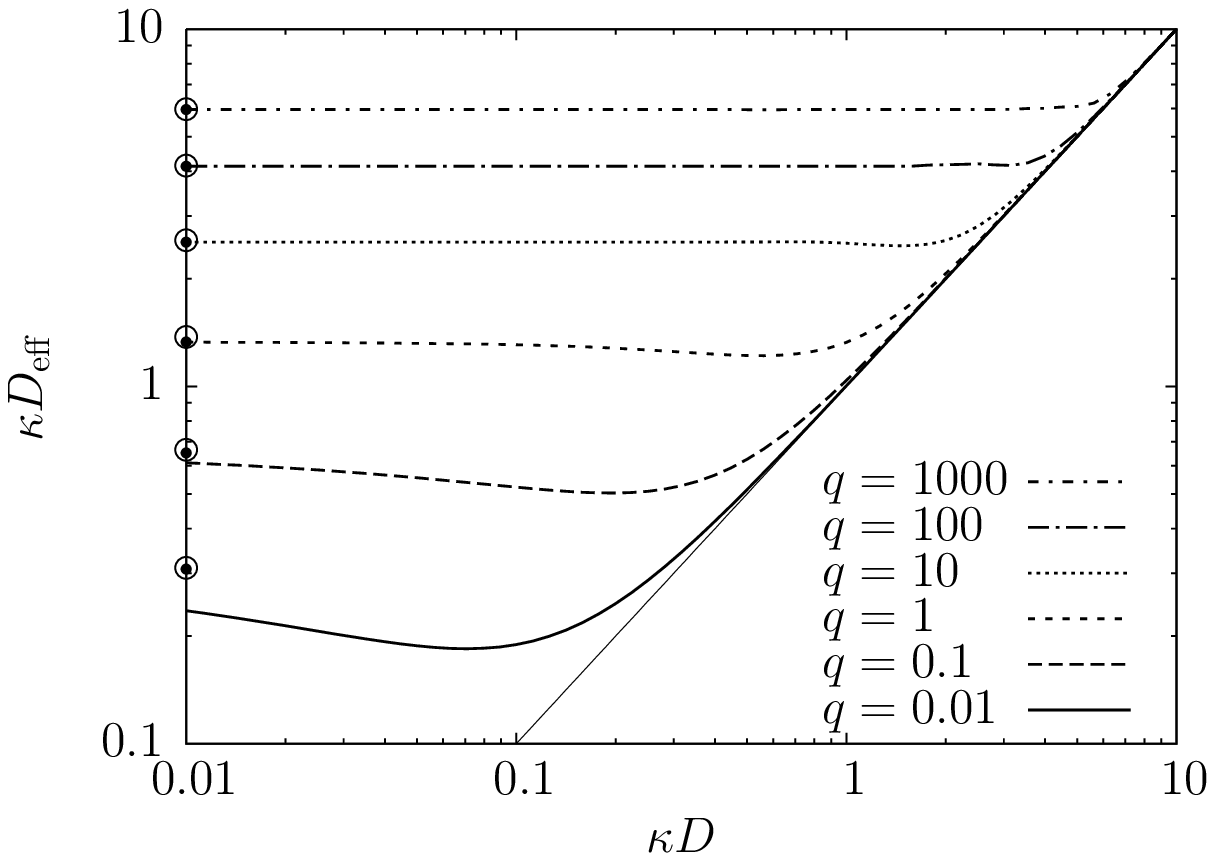}
\end{minipage}
\begin{minipage}{\miniwidth}
(b)
\includegraphics[width=\columnwidth]{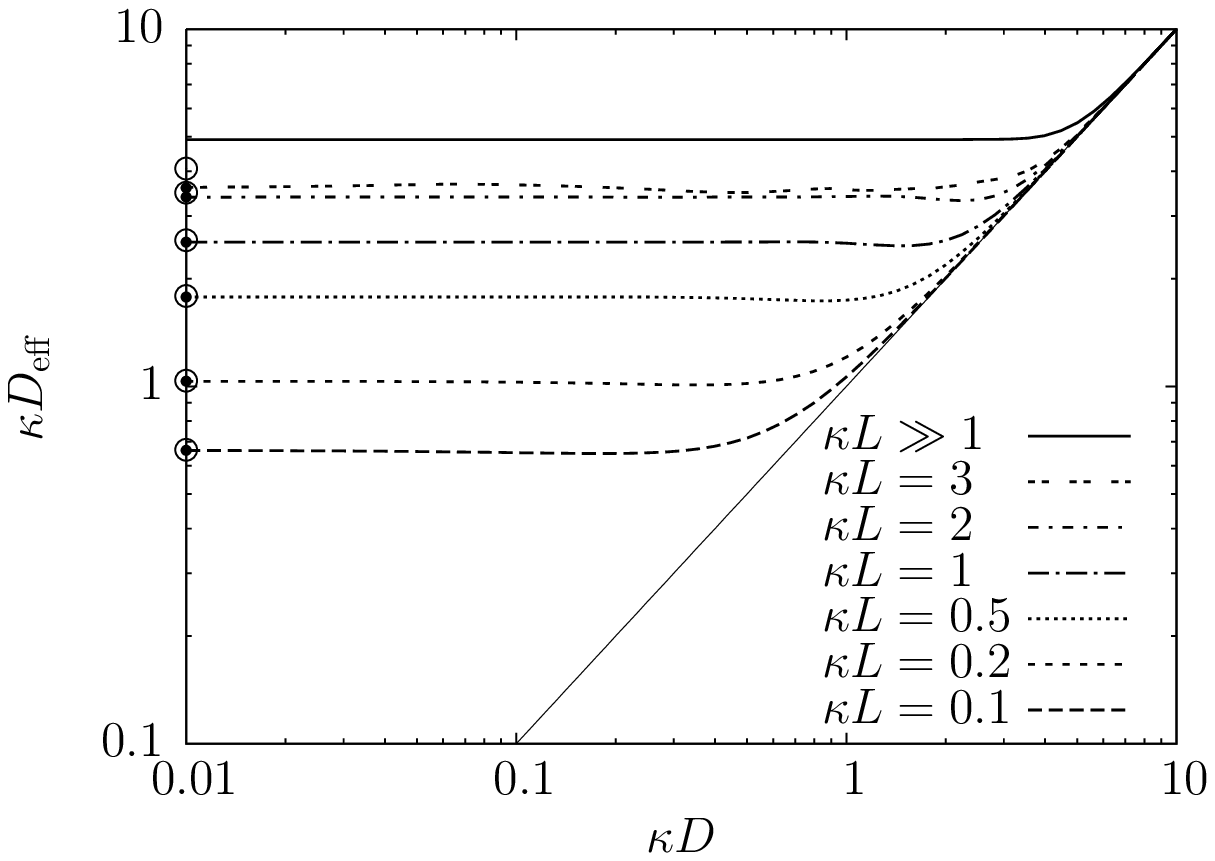}
\end{minipage}
\caption{The effective diameter $\kappa D_{\rm eff}$ as a function
of the rod diameter $\kappa D$ for (a) $\kappa L=1$ and different
values for the charge parameter $q$; (b) $q=10$ and different
values for the rod length $\kappa L$. The rod dimensions are
scaled by the screening length $\kappa^{-1}$. The thin solid line
is a guide to the eye, representing the hard-core limit $D_{\rm
eff}=D$. The small solid circles give the values for $\kappa D=0$
from the numerical calculations. The larger open circles are
obtained by the approximation given in Eq.~\eqref{eq:DLVO}.}
\label{fig:deff-d}
\end{figure*}

\begin{figure}[b]
\includegraphics[width=\figwidth]{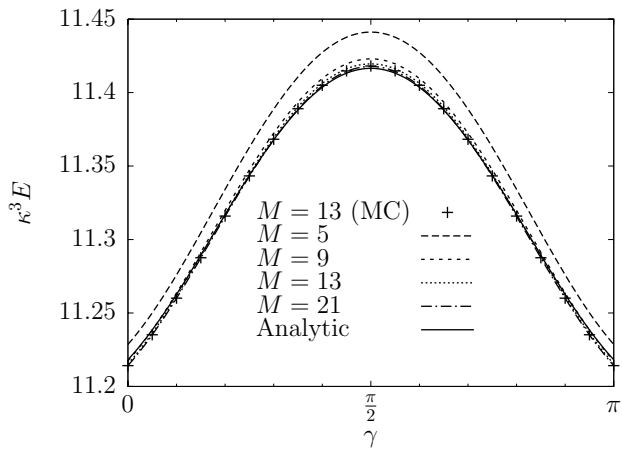}
\caption{The effective excluded volume $\kappa^{3}E$, as a
function of the angle $\gamma$ between the rod orientations,
calculated using different numerical schemes (see text), involving
$M$ discrete charges, Monte-Carlo (MC) integration, and the
present analytic approach. We used the parameter values $\kappa
L=1$, $\kappa D=0.25$ (such that $L/D=4$) and
$q=1$.}\label{fig:valid}
\end{figure}

In order to assess the accuracy of our calculations, we compare
some of the results of our calculations with those obtained from
more extensive numerical integration schemes. One is given by the
same spatial integration scheme as before, but with the
(effective) line-charge density replaced by a discrete charge
distribution. The rod charge is represented by an odd number of
charge units ($M=2N+1$) distributed evenly on the cylinder axis,
where one unit is always located on the center of the axis, and
two units are always located on the two end points of the axis.
The latter are of magnitude $e\lambda L/(4N)$, while all others
are of magnitude $e\lambda L/(2N)$. This ensures that the total
charge is $e\lambda L$ and the continuum limit
$N\rightarrow\infty$ yields the correct homogeneous line charge.
The other scheme uses the same discrete charge density as
described above, but uses a Monte-Carlo (MC) scheme to perform the
integration. This scheme is denoted by the plusses in
Fig.~\ref{fig:valid}. The agreement between the results obtained
from the different schemes, as shown in Fig.~\ref{fig:valid}, is
excellent for $M\geq 13$, particularly when considering that the
shape of the effective excluded volume differs significantly from
the hard-core case for these parameters. Therefore, we conclude
that our calculation correctly predicts the angular dependence of
the effective excluded volume of short charged rods.

\begin{figure*}[t]
\begin{minipage}{\miniwidth}
(a)
\includegraphics[width=\columnwidth]{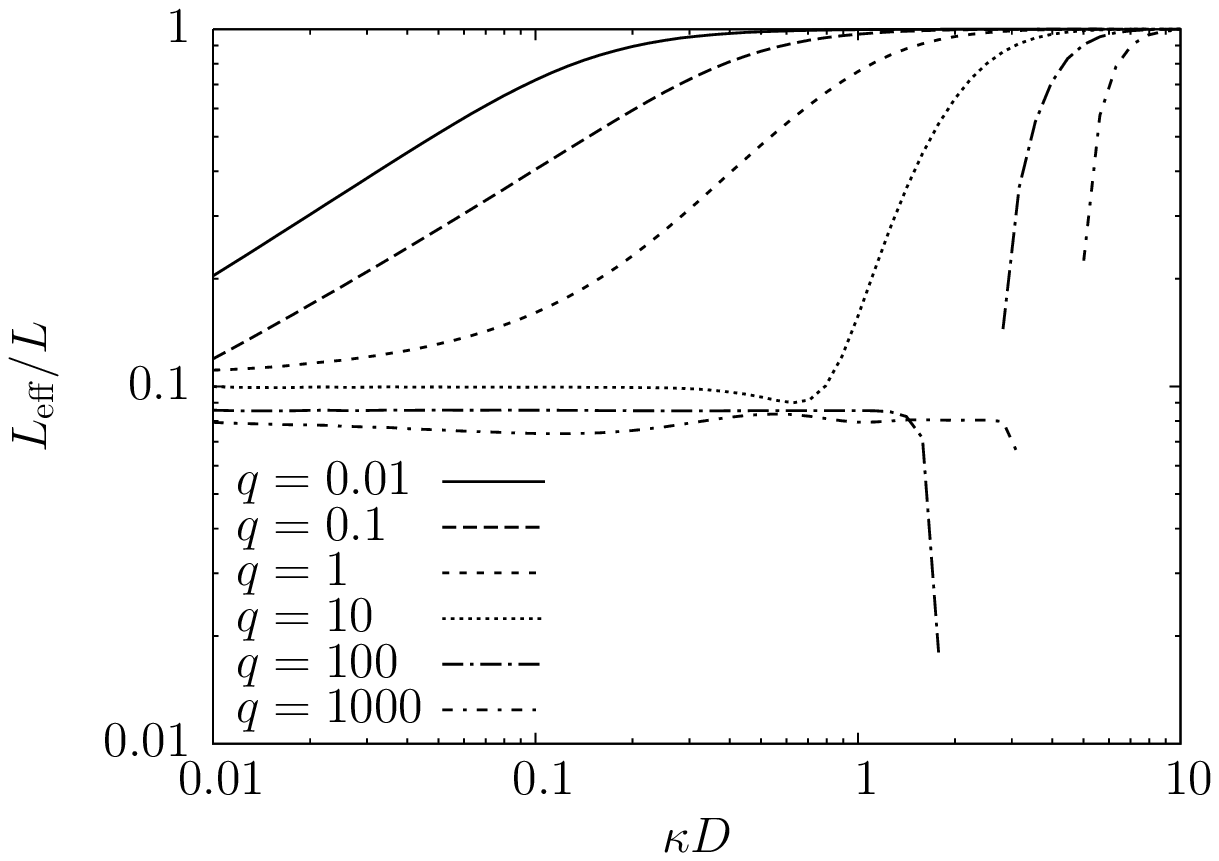}
\end{minipage}
\begin{minipage}{\miniwidth}
(b)
\includegraphics[width=\columnwidth]{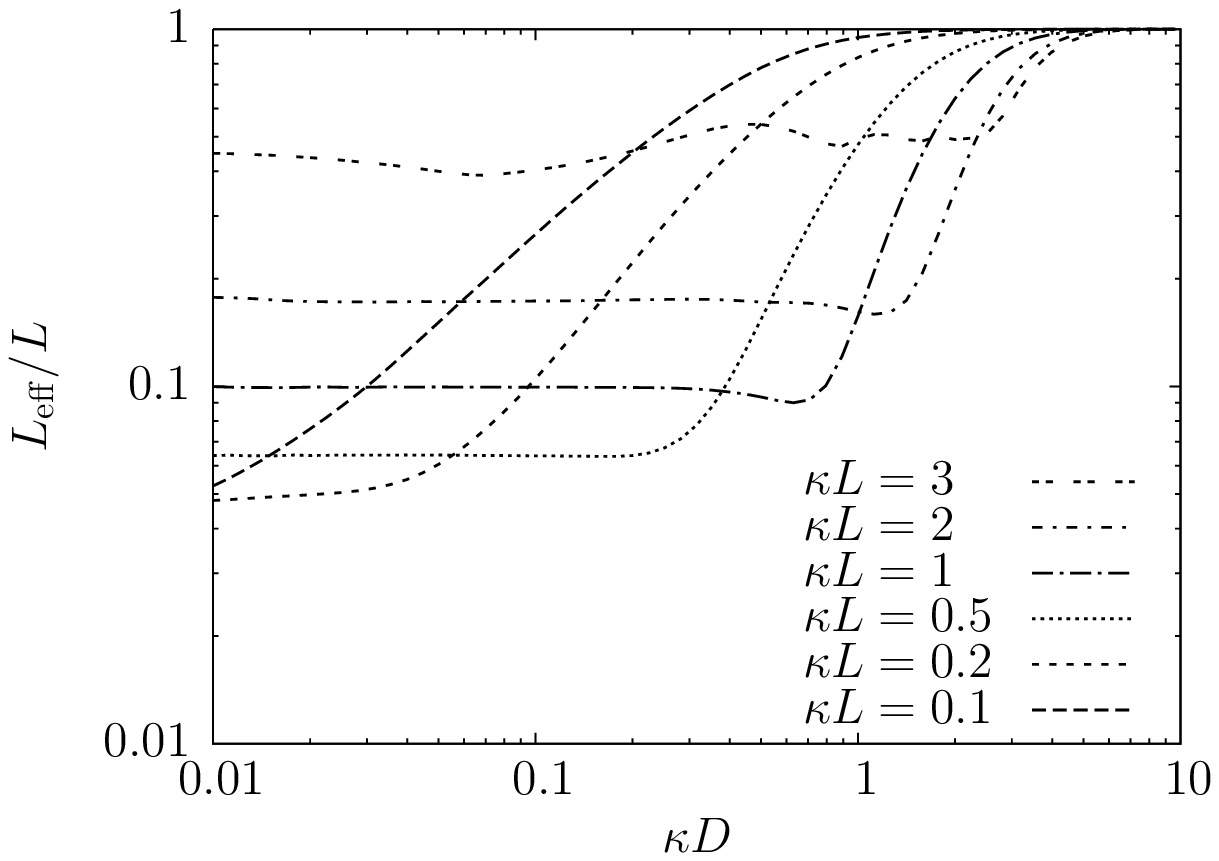}
\end{minipage}
\caption{The effective length $L_{\rm eff}/L$ as a function of the
rod diameter $\kappa D$ for (a) $\kappa L=1$ and different values
for the charge parameter $q$; (b) $q=10$ and different values for
the rod length $\kappa L$. The rod diameter is scaled by the
screening length $\kappa^{-1}$ and the effective length is scaled
by the rod length $L$.} \label{fig:leff-d}
\end{figure*}

In the previous section, we have shown that the angular dependence
of the effective excluded volume can be used to calculate the
effective rod dimensions $L_{\rm eff}$ and $D_{\rm eff}$|from the
values of $E_{\parallel}$ and $E_{\perp}$|by applying
Eqs.~\eqref{eq:deff}, \eqref{eq:ldeff}, and \eqref{eq:delta}.
Fig.~\ref{fig:deff-d}(a) shows the numerically calculated
effective diameter as a function of the real diameter for $\kappa
L=1$ and a range of charge parameters $q$.
Fig.~\ref{fig:deff-d}(b) shows the same function, but then for
$q=10$ and a range of rod lengths $\kappa L$. Note that all
(effective) rod dimensions are expressed in units of the screening
length. In Fig.~\ref{fig:deff-d}(b) the needle limit $\kappa L\gg
1$, given by Eq.~\eqref{eq:stroob}, is plotted for comparison.
Both graphs clearly reveal two regimes
\begin{equation}
D_{\rm eff} \simeq \left\{\begin{array}{cl} D_{\rm e} &\mbox{for }
D\ll D_{\rm e},\\\\
D &\mbox{for } D\gg D_{\rm e}.
\end{array}\right.
\end{equation}
These can be identified as an electrostatic regime at small
$\kappa D$ (weak screening) and a hard-core regime at high enough
$\kappa D$ (strong screening). In the hard-core regime, the
effective diameter equals the hard-core diameter, while in the
(weakly screened) electrostatic regime the effective diameter
saturates to a plateau value $D_{\rm e}$. This electrostatic
effective diameter depends on $q$ and $\kappa L$, and increases
with increasing $q$ and $\kappa L$. Also, it is (much) larger than
the hard-core diameter due to the (strong) rod-rod repulsions.
Values of the electrostatic effective diameter are included in
Fig.\ref{fig:deff-d}, where the small solid circles represent
values obtained from numerical calculations for $\kappa D=0$. The
larger open circles represent the following simple approximation
for $D_{\rm e}$.

In the short-rod limit, we can treat the double layer around the
rod as spherically symmetric, with an effective point charge
$e\lambda L$ in the center, such that also the pair potential is
spherically symmetric. This gives $\Delta=0$, and hence from
Eq.~\eqref{eq:deff}, for large enough $q$ (or small enough $\kappa
D$), we obtain the electrostatic effective diameter from the
simple expression
\begin{equation}\label{eq:DLVO}
\kappa D_{\rm e} \simeq \left[3\int_{0}^{\infty}{\rm
d}x\,x^{2}\left(1-\exp\left[-q\kappa^{2}L^{2}\frac{\exp[-x]}{x}\right]\right)\right]^{\frac{1}{3}}.
\end{equation}
This approximation is given in Fig.~\ref{fig:deff-d} by the larger
open circles. Both graphs show good agreement for $\kappa L\leq 1$
and all values for $q$. Fig.~\ref{fig:deff-d}(b) also shows that
the regime $\kappa L\leq 2$|which is reliably accessible with our
truncated numerical scheme|evolves smoothly to the needle-limit
$\kappa L\gg 1$ of \citet{Stroobants}. The curve for $\kappa L=3$
shows some signatures of the numerical instabilities we encounter
for larger $\kappa L$.

In a similar fashion we can also study the effective length
$L_{\rm eff}$ of the rods. Fig.~\ref{fig:leff-d}(a) shows results
of numerical calculations of the effective rod length for $\kappa
L=1$ and a range of charge parameters $q$.
Fig.~\ref{fig:leff-d}(b) is the result for $q=10$ and a range of
rod lengths $\kappa L$. The rod dimensions are expressed in units
of the Debye length, whereas $L_{\rm eff}$ is expressed in units
of the hard-core length. We distinguish again two asymptotic
regimes, the strong screening (hard core) regime $\kappa D\gg 1$
where $L_{\rm eff}=L$, and the weak-screening (electrostatic)
regime $\kappa D\ll 1$ where $L_{\rm eff}$ reaches a plateau value
that depends on $q$ and $\kappa L$. Note also that $L_{\rm eff}<L$
which is perhaps unexpected at first sight. Naively, one could
expect the effective length to increase with increasing effective
excluded volume. However, as \citet{Sato} pointed out, the
effective length decreases with increasing rod charge density
because of end effects. Thus, the increase of the effective
excluded volume|due to the increase of the rod charge density|is
purely caused by the increase of the effective diameter. Moreover,
this increase balances the decreasing in effective length such
that the total effective particle length $L_{\rm eff}+D_{\rm eff}$
does increase with increasing rod charge density. Inspection of
Fig.~\ref{fig:leff-d}(a) also reveals numerical (convergence)
problems for $q\geq 100$ at $\kappa D\gtrsim 1$, where $\kappa
L_{\rm eff}$ sharply drops and rises before reaching the hard-core
limit $L_{\rm eff}=L$. This is in fact only a minor problem in
practice, as it only occurs in the regime where $L_{\rm
eff}/D_{\rm eff}\lesssim 0.1$. There, the anisotropic contribution
to the effective excluded volume is much smaller than the
isotropic part. Upon approach of the needle-limit $\kappa L\gg 1$,
see Fig.~\ref{fig:leff-d}(b), we find that $L_{\rm eff}$
approaches $L$ for all values of $\kappa D$, as expected.

\section{Phase behaviour}
\begin{figure}[t]
\includegraphics[width=\figwidth]{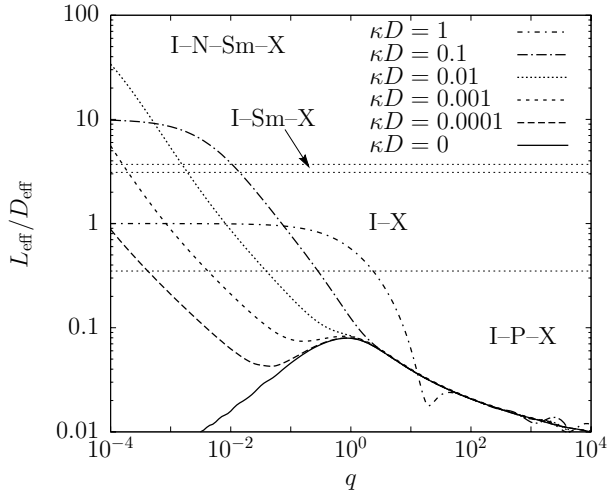} \caption{The effective
aspect ratio $L_{\rm eff}/D_{\rm eff}$ as a function of the charge
parameter $q$ for $\kappa L=1$ and different values for the rod
diameter $\kappa D$. Different|possibly coexisting|phases are
associated with a certain range of (effective) aspect ratios. See
the text for an explanation of the abbreviations and the boundary
values.} \label{fig:ldeff}
\end{figure}

We have determined the effective length and diameter of charged
rods, by mapping their orientation-dependent second virial
coefficient onto that of effective hard rods. Subsequently, we
also study the effective length-to-diameter ratio $L_{\rm
eff}/D_{\rm eff}$. In Fig.~\ref{fig:ldeff} we show this effective
aspect ratio as a function of the rod charge for $\kappa L=1$ and
a range of rod diameters $\kappa D$. All curves with $\kappa D>0$
essentially decrease from their maximum value|the hard-core aspect
ratio $L/D$|towards the curve given by $\kappa D=0$. This
indicates that the effective dimensions of charged rods become
independent of the hard-core diameter for large charge parameters,
where we enter the electrostatic regime. Also, since the effective
aspect ratio for $\kappa D=0$ is a decreasing function for large
$q$, we see that the charged rods essentially behave like charged
spheres upon increasing the charge above a certain value.

Moreover, Fig.~\ref{fig:ldeff} reveals a local maximum for very
small $\kappa D$, in the regime where $q\simeq 1$. This effect can
be understood by considering the electrostatic regime for small
charge parameters $q$. Eq.~\eqref{eq:ldeff} shows that the
effective aspect ratio is governed by the dimensionless anisotropy
parameter $\Delta$, which is defined in Eq.~\eqref{eq:delta}. In
the electrostatic regime, this anisotropy can be shown|up to first
order|to be proportional to $q$. The reason for this is that the
linear approximation of the effective excluded volume is
orientation independent \citep{Chen-Koch}. Therefore, the
difference between the isotropically-averaged and parallel values
is of second order in $q$, whereas the parallel value itself is of
first order. The effective aspect ratio is of order
$\sqrt{\Delta}$, and thus increases as $\sqrt{q}$. Conversely, for
$q\gtrsim 1$ the effective length is more or less constant, and
the effective aspect ratio decreases again due to the increase of
the effective diameter.

The horizontal dotted lines in Fig.~\ref{fig:ldeff} indicate the
crossover values (0.35, 3.5, and 3.7) for regimes with different
phase sequences. The values for these aspect ratios are taken from
simulation results of hard-spherocylinder systems by
\citet{Bolhuis}. These simulations consist of explicit free-energy
calculations of coexisting phases, where the most dilute phase is
always given by an isotropic fluid (I), and the most dense phase
by a fully ordered crystal (X). Depending on the aspect ratio,
different phases were found in between these two phases. For
aspect ratios exceeding $\sim 3.7$ the phase sequence I--N--Sm--X
was found upon increasing the density. Here, the N and Sm denote
the nematic and smectic-A liquid crystalline phases, respectively.
Somewhat shorter rods, with an aspect ratio in the narrow regime
between $\sim 3.5$ and $\sim 3.7$, can still form a smectic-A but
no longer a nematic phase, and hence have a phase sequence
I--Sm--X. Even shorter hard rods, with an aspect ratio in between
$\sim 0.35$ and $\sim 3.5$ cannot form a thermodynamically stable
smectic-A phase, and thus crystallize directly into a fully
ordered crystal from the isotropic fluid, yielding a phase
sequence I--X. Very short hard rods, with an aspect ratio smaller
than $\sim 0.35$, exhibit a plastic (P) crystal phase, such that
the phase sequence is I--P--X. The plastic crystal phase is
characterized by orientational disorder, but has translational
order as in a crystal phase \citep{Bolhuis,Vega}. This regime
arises naturally in the case that $\kappa L$ is small. Then, such
a crystal forms because of the essentially isotropic long-range
repulsive interactions, but the competition with entropic effects
prevents the rods from aligning.

We use the mapping of the charged-rod system onto the effective
hard-rod system to give an indication of the phase sequence of
systems of charged rods as a function of the parameters $\kappa
L$, $\kappa D$ (or $L/D$), and $q$. For instance, from the curve
for $\kappa D=1$ in Fig.~\ref{fig:ldeff}, we see that the
effective aspect ratio never exceeds unity for any $q$. This
excludes the possibility of a nematic or smectic-A liquid crystal
phase. The curve starts off at its maximum (in the limit where
$q\rightarrow 0$), where the effective aspect ratio equals the
hard-core aspect ratio $L/D=\kappa L=1$. It crosses the value
$L_{\rm eff}/D_{\rm eff}=0.35$ at $q\approx 2.35$, such that a
sufficiently large rod charge density allows for a plastic crystal
phase. Similarly, for $\kappa D=0.1$ (which corresponds to
$L/D=10$), we find all four phase sequences upon increasing $q$.

\begin{figure*}[ht]
\begin{minipage}{\miniwidth}
(a)
\includegraphics[width=\columnwidth]{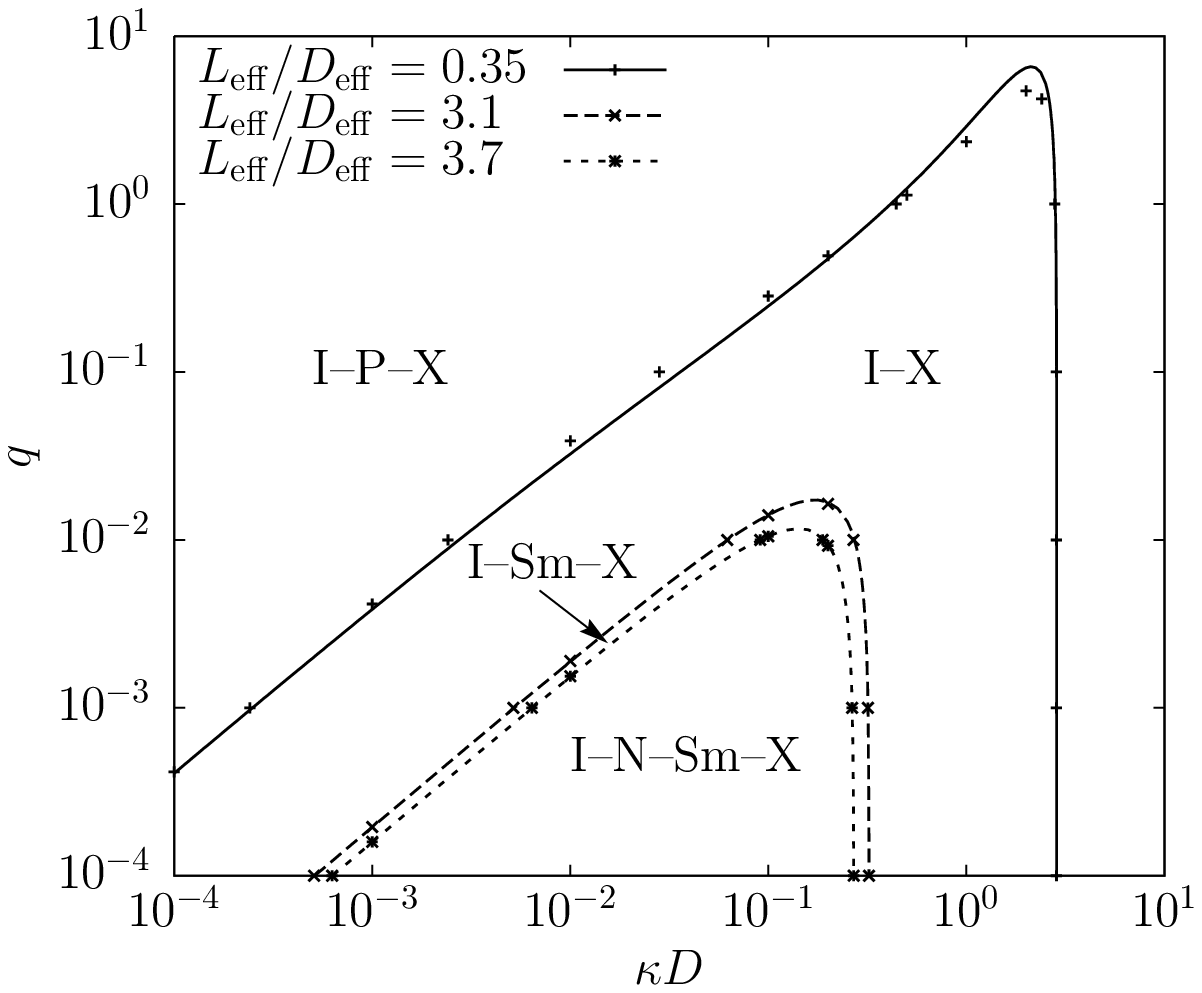}
\end{minipage}
\begin{minipage}{\miniwidth}
(b)
\includegraphics[width=\columnwidth]{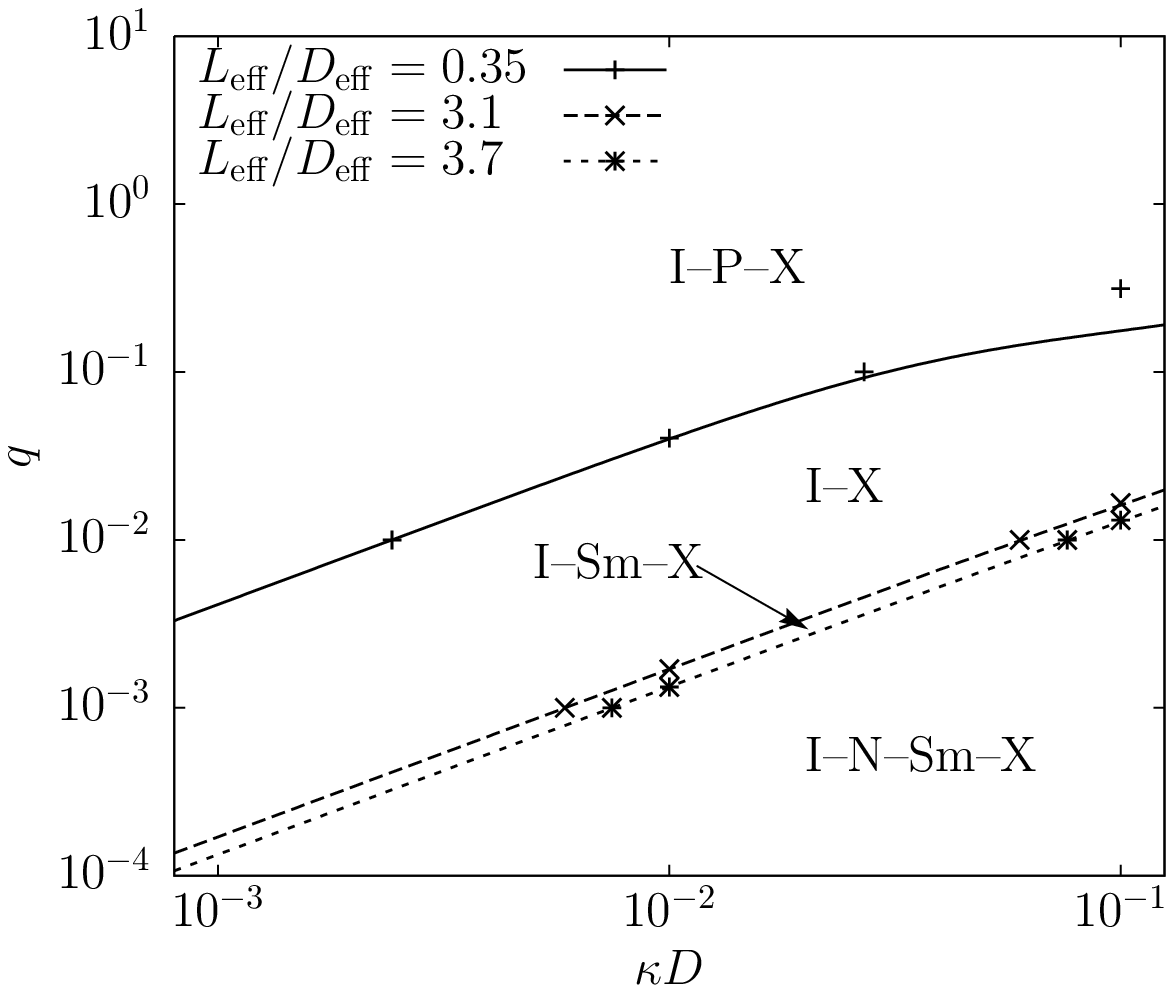}
\end{minipage}
\caption{Boundary lines for given values for the effective aspect
ratio $L_{\rm eff}/D_{\rm eff}$. See the text for an explanation
of the abbreviations of the different regime labels. The points
are results of the numerical calculations, and the lines are given
by a simplified theory. We fix (a) $\kappa L=1$, and (b) $L/D=20$,
respectively.} \label{fig:ldcont}
\end{figure*}

By determining the intersections of the effective aspect ratio
with the crossover values of the hard-rod system, we construct
``phase diagrams'' indicating the different regimes. In
Fig.~\ref{fig:ldcont} we present two examples of such diagrams in
the plane spanned by $q$ and $\kappa D$. In
Fig.~\ref{fig:ldcont}(a), we fix $\kappa L=1$, such that the
horizontal axis could read $D/L$ as well. In
Fig.~\ref{fig:ldcont}(b) we fix $L/D=20$, such that the change in
$\kappa D$ physically corresponds to a change in salt
concentration (while keeping the particle dimensions fixed). The
symbols denote the crossover values for the effective aspect ratio
as determined from our numerical data (such as presented in
Fig.~\ref{fig:ldeff}). The lines are based on an approximate
theoretical model to be discussed in section \ref{sec:theory}.

Both diagrams in Fig.~\ref{fig:ldcont} show that rods with
sufficiently high surface charge density always show the I--P--X
sequence. This is due to the essentially spherical nature of the
effective shape of highly charged rods. The limit of uncharged
rods is determined by the hard-core sequence that corresponds to
$L/D$. The I--N--Sm--X regime at fixed $\kappa L$ in
Fig.~\ref{fig:ldcont}(a) is completely bounded. First, by a
hard-core regime when $\kappa D\gtrsim 0.27$, where the liquid
crystal phases cannot exist even for $q=0$ because $L/D\lesssim
3.7$. Second, by an electrostatic regime in the weak-screening
limit of small $\kappa D$, where the rods effectively behave as
spheres since $D_{\rm eff}\gg L_{\rm eff}$. Conversely, the trends
displayed for fixed $L/D$ in Fig.~\ref{fig:ldcont}(b) are
monotonic, with an I--N--Sm--X regime that extends to higher $q$
with increasing $\kappa D$.

\begin{figure*}[t]
\begin{minipage}{\miniwidth}
(a)
\includegraphics[width=\columnwidth]{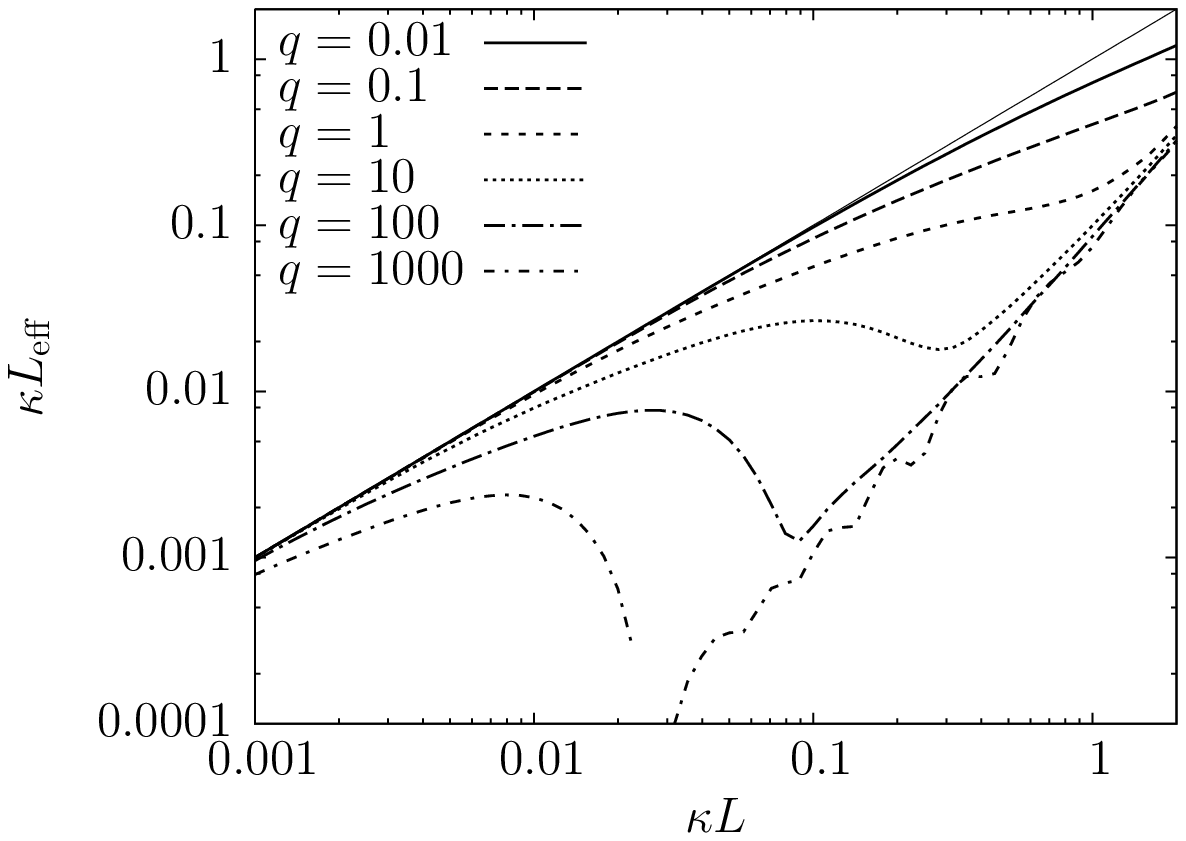}
\end{minipage}
\begin{minipage}{\miniwidth}
(b)
\includegraphics[width=\columnwidth]{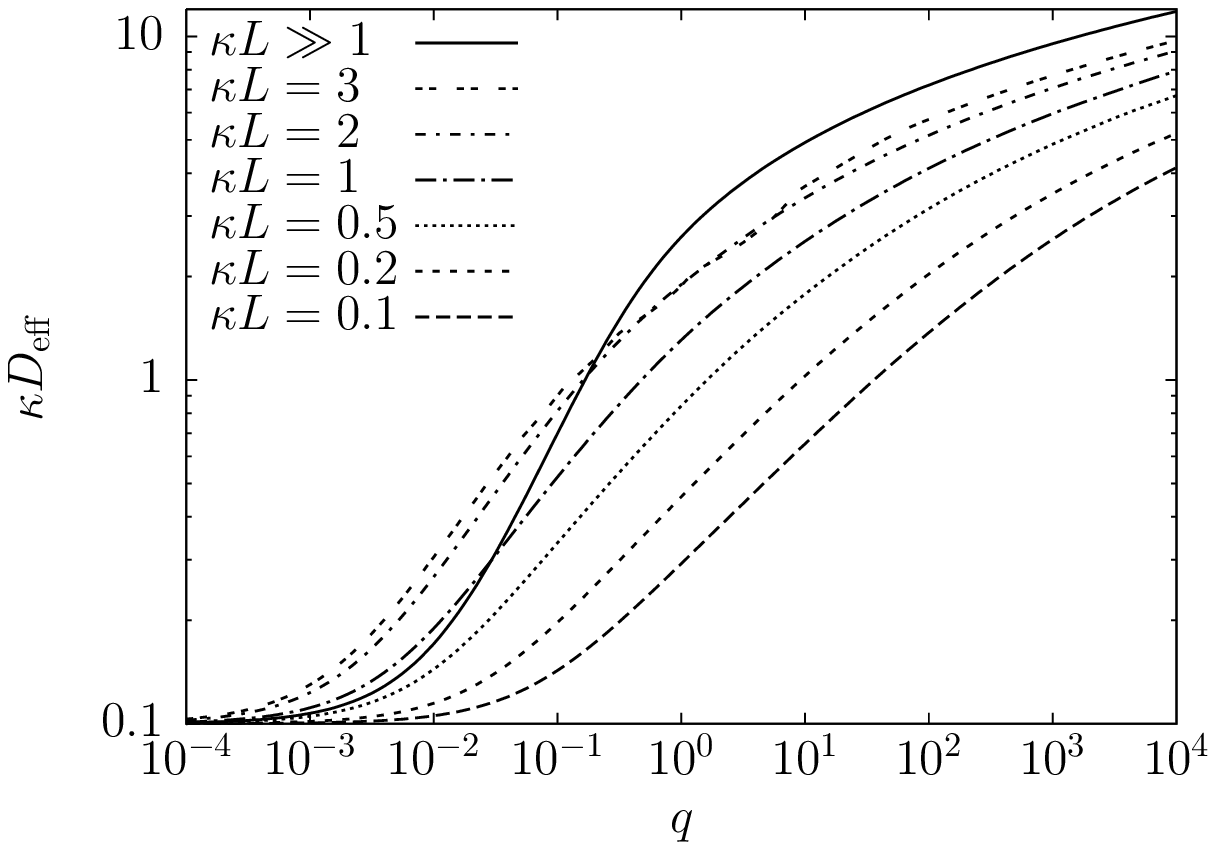}
\end{minipage}
\caption{(a) The effective length $\kappa L_{\rm eff}$ as a
function of the rod length $\kappa L$ for $\kappa D=0.1$ and
different values for the charge parameter $q$. The thin solid line
represents the needle or hard-core limit, where $L_{\rm eff}\simeq
L$. (b) The effective diameter $\kappa D_{\rm eff}$ as a function
of the rod charge parameter $q$ for $\kappa D=0.1$ and different
values for the rod length $\kappa L$. The (effective) rod
dimensions are scaled by the screening length
$\kappa^{-1}$.}\label{fig:dim-kd01}
\end{figure*}

\section{A Simpler Model} \label{sec:theory}
For small values of the effective surface-charge density, we found
that the electrostatic contribution to the effective excluded
volume is essentially isotropic in nature. This means that the
anisotropic effects are primarily due to the hard-core anisotropy
(as apparent from Fig.~\ref{fig:excl-vol}), such that
\begin{equation}
L_{\rm eff}^{2}D_{\rm eff} \simeq L^{2}D.
\end{equation}
On this basis, we propose here a simple model, which turns out to
describe our numerical findings with remarkable accuracy. This
model introduces a ``spherical approximation'' of the
electrostatic contribution to the effective excluded volume, which
involves the orientation-dependent diameter $\bar{D}(\gamma)$. The
volume of a sphere of this diameter is equal to the hard-core
excluded volume of a pair of rods
\begin{equation}\label{eq:dbar}
\frac{4\pi}{3}\bar{D}(\gamma)^{3} = \frac{4\pi}{3}D^{3} + 2\pi
LD^{2} + 2L^{2}D\sin\gamma.
\end{equation}
We approximate the effective excluded volume by the value for a
charged sphere of diameter $\bar{D}(\gamma)$ and an effective
surface charge that equals the total amount of effective charge on
the rods
\begin{align}\label{eq:ebar}
&\bar{E}(\gamma) = \frac{4\pi}{3}\bar{D}(\gamma)^{3}\nonumber\\
&{} + 4\pi\int_{\bar{D}(\gamma)}^{\infty}{\rm
d}r\,r^{2}\left(1-\exp\left[-q\kappa^{2}L^{2}\frac{\exp[-\kappa
r]}{\kappa r}\right]\right).
\end{align}
Note that the only orientation dependence of the electrostatic
contribution to this effective excluded volume (i.e.~the second
term) comes from the integral boundary $\bar{D}(\gamma)$. To
calculate the effective dimensions, we only need the parallel and
isotropically averaged values of the effective excluded volume. In
the parallel case ($\gamma=0$) this value is readily calculated
\begin{align}\label{eq:ebarp}
&\bar{E}_{\parallel} = \frac{4\pi}{3}\bar{D}_{\parallel}^{3}\nonumber\\
&{} + 4\pi\int_{\bar{D}_{\parallel}}^{\infty}{\rm
d}r\,r^{2}\left(1-\exp\left[-q\kappa^{2}L^{2}\frac{\exp[-\kappa
r]}{\kappa r}\right]\right),
\end{align}
where
\begin{equation}
\frac{4\pi}{3}\bar{D}_{\parallel}^{3} = \frac{4\pi}{3}D^{3} + 2\pi
LD^{2}.
\end{equation}
The isotropically-averaged value can be calculated numerically by
using expression \eqref{eq:ebar}. However, we approximate it by
the value for a charged sphere of diameter $\bar{D}_{\rm iso}$
(using the same total effective charge), which is taken from the
isotropic average of the hard-core excluded volume
\begin{equation}
\frac{4\pi}{3}\bar{D}_{\rm iso}^{3} = \frac{4\pi}{3}D^{3} + 2\pi
LD^{2} + \frac{\pi}{2}L^{2}D.
\end{equation}
This approximation yields the simple expression
\begin{align}\label{eq:ebari}
&\bar{E}_{\rm iso} = \frac{4\pi}{3}\bar{D}_{\rm iso}^{3}\nonumber\\
&{} + 4\pi\int_{\bar{D}_{\rm iso}}^{\infty}{\rm
d}r\,r^{2}\left(1-\exp\left[-q\kappa^{2}L^{2}\frac{\exp[-\kappa
r]}{\kappa r}\right]\right).
\end{align}
With our explicit expressions \eqref{eq:ebarp} and
\eqref{eq:ebari}, we evaluate the effective dimensions from
Eqs.~\eqref{eq:deff} and \eqref{eq:ldeff} as before. The resulting
crossover values of the hard-rod system are shown by the curves in
Fig.~\ref{fig:ldcont}, and are in very good agreement with the
numerical calculations (denoted by the symbols). The key to this
remarkable accuracy lies in the fact that the anisotropic
electrostatic contributions are relatively unimportant, because
the rod length is small with respect to the screening length
(i.e.~$\kappa L\leq 2$). Thus, our simple model accounts for the
hard-core anisotropy correctly, as well as for the isotropic
electrostatic contribution.

In a sense, this theoretical description can be viewed as a kind
of perturbation theory, where we expand the pair potential as a
function of $\kappa L$. The hard-core repulsion represents the
zero-th order. The lowest-order contribution to $V_{\rm e}$ is
quadratic in $\kappa L$ and independent of rod orientations. Also,
it happens to correspond to the interaction potential of two point
charges $e\lambda L$. If we plug this approximation of
$V(\mathbf{r};\hat{\omega},\hat{\omega}')$ into the expression of
the effective excluded volume (given by Eq.~\eqref{eq:E}), we
obtain an expression where the integral boundary $\bar{D}$ is
still a function of both the angle between the rod orientations
and the direction of the center-to-center separation vector
$\mathbf{r}$. In fact, it is given by the distance where the rods
touch, given a certain orientational configuration. By setting
this overlap diameter to a value that is independent of the
orientation of $\mathbf{r}$, but still respects the total
hard-core excluded volume, we effectively neglect its dependence
on $\kappa L$. This choice is justified by the fact that (for
small $\kappa L$) the size of the double layer around the
particles is larger than the variations in the overlap diameter
$\bar{D}$. That is why our simple theoretical description can be
interpreted as a perturbation theory of a hard-rod reference
system with an (almost) isotropic electrostatic contribution.
Unfortunately, it completely fails to describe the anisotropic
effects in the electrostatic regime. In this regime the
anisotropic details of the electrostatic contributions do become
important compared to the hard-core contributions.

\section{Discussion and Conclusion}
The numerical results presented in this paper give access to a
part of the parameter space where there is a large difference
between the effective length and the real length. In this regime,
one cannot hope that the theory of \citet{Stroobants} gives any
accurate results, as this is based on the needle limit where
$L_{\rm eff}\simeq L$. The perturbation theory of
\citet{Chen-Koch} breaks down for most of our parameter values.
This is because it is based on small charges, and thus fails to
describe the effect of large rod surface-charge densities. Also,
this theory is not accurate for large differences between the
effective and hard-core diameter.

In Fig.~\ref{fig:dim-kd01}(a), we show results of numerical
calculations of the effective rod length as a function of the
hard-core length, for $\kappa D=0.1$. Note that again the
effective length is always smaller than (or equal to) the
hard-core length. Also, in accordance with the results from
Fig.~\ref{fig:leff-d}, there is a hard-core regime for small
values of the charge parameter $q$, as well as for small values of
the rod length $\kappa L$, for which the total amount of effective
rod charge is small. On the other hand, there is an electrostatic
regime. In Fig.~\ref{fig:leff-d}, this was shown to be the case
for decreasing values of $\kappa D$, where the plateau value
(i.e.~the electrostatic length) depends on $q$ and $\kappa L$.
However, from Fig.~\ref{fig:dim-kd01}(a), it can be seen that this
electrostatic length depends mostly on the rod length $\kappa L$,
and not really on the charge parameter $q$, as long as either $q$
or $\kappa L$ is large enough. Furthermore, the effective length
is ``wedged'' in between the electrostatic length and the
hard-core length, where the electrostatic length approaches the
hard-core length in the needle limit ($\kappa L\gg 1$).
Unfortunately, there is no analytic theory yet that describes our
numerical results for this electrostatic length as a function of
$\kappa L$. Therefore, it would be worthwhile to gain new insight
in the effect of electrostatics on the effective rod length for
intermediate $\kappa L$|neglecting hard-core interactions|in the
case of large rod charges. Additionally,
Fig.~\ref{fig:dim-kd01}(b) shows results of numerical calculations
of the effective diameter as a function of the charge parameter
$q$. For $q\gtrsim 1$, there is a smooth transition to the
theoretical needle limit of Ref.~\citep{Stroobants}, where $\kappa
L\gg 1$. Conversely, this is not the case for $q\lesssim 1$, due
to the fact that the approximations leading to
Eq.~\eqref{eq:stroob} do not give the correct effective excluded
volume for small values of $q$ and (nearly) parallel rods. More
investigations need to be made into this regime.

In conclusion, we have numerically studied the second virial
coefficient of short charged rods dispersed in an electrolyte,
presuming pairwise screened-Coulomb interactions between the
line-charge segments of the rods. The control parameters of
interest are the hard-core length $L$ and diameter $D$, the Debye
screening length of the medium $\kappa^{-1}$, and the charge
parameter $q$. The main resulting quantities are the effective
diameter $D_{\rm eff}$ and length $L_{\rm eff}$ of the rods.  By a
mapping onto an effective hard-core system|for which the sequence
of phases between the dilute isotropic phase and the dense
crystalline phase is known for all aspect ratios|we predict the
relations between control parameters and the expected phase
sequence explicitly. We have also constructed a simplified model,
based on the diameter $\bar{D}(\gamma)$ of Eq.~\eqref{eq:dbar},
which reproduces the numerical results accurately at the expense
of much less computational effort. This model is particularly
successful in the regime of large effective aspect ratios ($L_{\rm
eff}/D_{\rm eff}>1$) and small ratios of the rod length to the
screening length ($\kappa L<1$).

An important result of this work is that highly charged short rods
at low salt concentrations (i.e.~at strong Coulomb couplings) have
a strong tendency to form plastic crystals upon compression. The
plasticity stems from the large effective diameter, which make the
rods behave essentially as inflated repulsive spheres with only
small nonspherical interactions that are too weak to cause
orientational ordering in the crystalline phase. This finding
could be important in the study of silica or gold nanorods, that
have reasonably large hard-core aspect ratio (like $L/D\simeq 5$).
Here, liquid crystalline phases could be expected, but only if the
charge on the rods is small enough.

\acknowledgements
It is a pleasure to thank Ahmet Demir\"{o}rs for
explaining his preliminary experimental results on charged
dumbbells.

\appendix
\section{The pair interaction of two charged rods}
The pair interaction of two charged rods is given by
Eq.~\eqref{eq:electro}, where we assume that the electrostatic
interaction is determined by integrating over pairs of effective
line-charge elements interacting with the screened Coulomb
potential. The distance between these pairs is given by a
superposition of the relative position of the rods and the
combination of the position of the line elements along both rods.
Since the integral in Eq.~\eqref{eq:electro} cannot be calculated
analytically, we try to simplify the calculation. By expanding the
integrand in spherical harmonics, we obtain terms that factorize
into two functions of the respective positions
\begin{align}
\frac{\exp[-|\mathbf{r}-\mathbf{s}|]}{|\mathbf{r}-\mathbf{s}|}
={}& \sum_{l=0}^{\infty}(2l+1)k_{l}(r)
P_{l}(\hat{\mathbf{r}}\cdot\hat{\mathbf{s}})
i_{l}(s)\quad\mbox{for }r>s,\nonumber\\
={}& 4\pi\sum_{l=0}^{\infty}\sum_{m=-l}^{+l}
k_{l}(r)Y_{l,m}(\hat{\mathbf{r}})i_{l}(s)Y_{l,m}^{*}(\hat{\mathbf{s}}),\label{eq:sph}
\end{align}
where $i_{l}$ and $k_{l}$ are the modified spherical Bessel
functions of the first and second kind, respectively. These
functions are given by
\begin{align}
i_{l}(x) ={}& \sqrt{\frac{\pi}{2x}}I_{l+\frac{1}{2}}(x),\\
k_{l}(x) ={}& \sqrt{\frac{2}{\pi x}}K_{l+\frac{1}{2}}(x),
\end{align}
where $I_{\nu}$ and $K_{\nu}$ are the modified (cylindrical)
Bessel functions of the first and second kind, respectively. The
Legendre polynomials $P_{l}$ are expanded into spherical harmonics
$Y_{l,m}$ using the famous addition theorem. We use the notation
where $r=|\mathbf{r}|$, and $\hat{\mathbf{r}}=\mathbf{r}/r$.
Finally, the asterisk ``$*$'' denotes complex conjugation. The
unit vector as given in the arguments of each of the spherical
harmonic functions should be interpreted as the two angles in
spherical coordinates with respect to an arbitrarily chosen
reference frame. Since the Legendre polynomials of the dot product
of the two orientations is independent of this choice, so is the
sum over $m$ of the product of the two spherical harmonics.

We note that one could consider rewriting the expression of the
pair potential in rotational invariants (as used in
Ref.~\citep{Graf:TMV}). These are functions of three orientations,
including a sum over $m$ of a product of three spherical harmonic
functions multiplied by Clebsch-Gordon coefficients. They form a
complete set of orthogonal functions dependent only on the
relative orientations of $\hat{\mathbf{r}}$,
$\hat{\mathbf{\omega}}$, and $\hat{\mathbf{\omega}}'$ with respect
to each other . However, it turns out that in our case these are
not really helpful. Alternatively, one could consider a
resummation of the expansion in spherical harmonics, such that
each term has a faster asymptotic decay than the previous term.
This is not the case here, since each Bessel function $k_{l}$ has
the same asymptotic decay as $k_{0}$ \citep{Ramirez}.

\section{Domains of integration}

\begin{figure}[ht]
\includegraphics[width=\figwidth]{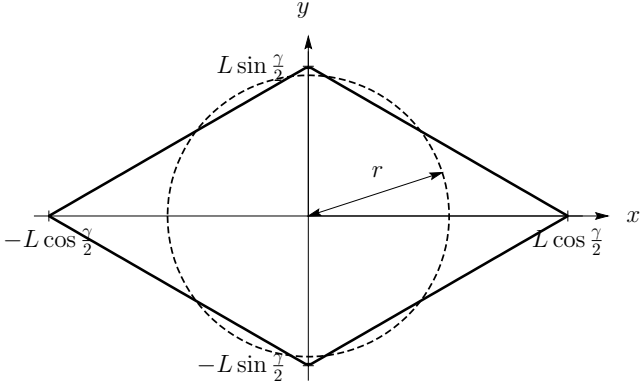}
\caption{Illustration of the domain of integration of the
superposition of the positions of two line elements. The dashed
circle of radius $r$ divides the parallelogram into two
domains.}\label{fig:paral}
\end{figure}

The integration over line elements of both rods in
Eq.~\eqref{eq:electro} is in fact an integration of the vector
$l\hat{\omega}-l'\hat{\omega}'$ over a parallelogram-shaped area
in the plane tangent to both rod orientations. This area is
illustrated in Fig.~\ref{fig:paral}. There is a straightforward
choice for the reference frame and a substitution of variables
\begin{align}
\hat{\omega} ={}& \left(\cos\frac{\gamma}{2},\sin\frac{\gamma}{2},0\right),\label{eq:frame}\\
\hat{\omega}' ={}& \left(\cos\frac{\gamma}{2},-\sin\frac{\gamma}{2},0\right),\\
l\hat{\omega}-l'\hat{\omega}' ={}&
(\rho\cos\varphi,\rho\sin\varphi,0)\label{eq:polar},
\end{align}
where $\gamma$ is the angle between the two rod orientations. The
polar coordinates $\rho$ and $\varphi$ describe the same plane as
$l$ and $l'$. The parallelogram can be cut up into four equivalent
pieces, keeping only the terms in the expansion \eqref{eq:sph}
where $l$ and $m$ are both even. The integral boundaries of the
first quadrant ($0\leq\varphi\leq\pi/2$) satisfy
\begin{equation}
0 \leq \rho \leq
\frac{L\sin\gamma}{2\sin\left(\varphi+\frac{\gamma}{2}\right)}.
\end{equation}

It is important to note that the functional form of the integrant
can vary as a function of $\rho$, because $k_l$ and $i_l$ switch
roles when $r<\rho$. We shall split the result of our expansion
into each order in $l$ and $m$, to be examined separately. We
write
\begin{align}
\beta V_{\rm e}(r,\theta,\phi;\hat{\omega},\hat{\omega}')&\nonumber\\
{}= \kappa l_{\rm B}\lambda^{2}
\underbrace{\sum_{l=0}^{\infty}\sum_{m=-l}^{+l}}_{l,m\;{\rm even}}
&\frac{(-1)^{\frac{l+m}{2}}(2l+1)(l-m)!}{2^{l}\left(\frac{l+m}{2}\right)!\left(\frac{l-m}{2}\right)!}\nonumber\\
&{}\times
\mathcal{A}_{l,m}(r;\gamma)P_{l,m}(\cos\theta)\cos(m\phi),\label{eq:elecsph}
\end{align}
where $P_{l,m}$ are the associated Legendre functions. We have
used that for $l$ and $m$ both even
\begin{align}
&\frac{Y_{l,m}(\theta,\phi) + Y_{l,-m}(\theta,\phi)}{2}\nonumber\\
&{}=
\sqrt{\frac{2l+1}{4\pi}\frac{(l-m)!}{(l+m)!}}\,P_{l,m}(\cos\theta)\cos(m\phi),
\end{align}
and
\begin{align}
&\frac{1}{2}\left(Y_{l,m}^{*}\left(\vartheta=\frac{\pi}{2},\varphi\right)
+
Y_{l,-m}^{*}\left(\vartheta=\frac{\pi}{2},\varphi\right)\right)\nonumber\\
&{}=
(-1)^{\frac{l+m}{2}}\sqrt{\frac{2l+1}{4\pi}}\frac{\sqrt{(l+m)!(l-m)!}}{2^{l}\left(\frac{l+m}{2}\right)!\left(\frac{l-m}{2}\right)!}\,\cos(m\varphi).
\end{align}
The integral $\mathcal{A}_{l,m}(r;\gamma)$ in
Eq.~\eqref{eq:elecsph} is given by
\begin{equation}\label{eq:inta}
\mathcal{A}_{l,m}(r;\gamma) =
\frac{4}{\sin\gamma}\int_{0}^{\frac{\pi}{2}}{\rm
d}\varphi\,\cos(m\varphi)\mathcal{B}_{l}(r;\varphi,\gamma),
\end{equation}
where
\begin{equation}\label{eq:multi}
\mathcal{B}_{l}(r;\varphi,\gamma) = k_{l}(\kappa r)
\int_{0}^{\frac{L\sin\gamma}{2\sin\left(\varphi+\frac{\gamma}{2}\right)}}
{\rm d}\rho\,\rho\,i_{l}(\kappa\rho)
\end{equation}
for
$r>\frac{L\sin\gamma}{2\sin\left(\varphi+\frac{\gamma}{2}\right)}$,
and
\begin{align}
\mathcal{B}_{l}(r;\varphi,\gamma) ={}& k_{l}(\kappa
r)\int_{0}^{r}{\rm
d}\rho\,\rho\,i_{l}(\kappa\rho)\nonumber\\
&{}+ i_{l}(\kappa r)
\int_{r}^{\frac{L\sin\gamma}{2\sin\left(\varphi+\frac{\gamma}{2}\right)}}
{\rm d}\rho\,\rho\,k_{l}(\kappa\rho)\label{eq:local}
\end{align}
for
$r<\frac{L\sin\gamma}{2\sin\left(\varphi+\frac{\gamma}{2}\right)}$.

Let us have another look at Fig.~\ref{fig:paral}. The dashed
circle indicates the value for which the variables $r$ and $s$ in
Eq.~\eqref{eq:sph} switch (in this case $s$ is replaced by
$\rho$). Consider the first quadrant (i.e.\ the upper right-hand
corner). Let us also assume $\gamma<\pi/2$. In the end, we will
calculate the effective excluded volume for $0<\gamma<\pi$, but
this expression is symmetric in $\gamma\leftrightarrow\pi-\gamma$
(due to up-down symmetry) so we need only the first half of this
interval. We describe the integral boundary for $\rho$ as a
function of $\varphi$ just as we describe the boundary of the
parallelogram by $\rho$ as a function of $\varphi$. However, the
integrand in $\mathcal{B}$ changes when the boundary of the
parallelogram intersects with the circle of radius $r$.
Therefore|depending on the value of $r$|we have one to three
domains for $\mathcal{B}$ as a function of $\varphi$
\begin{equation*}
\begin{array}{l}
\varphi\in\left[0,\frac{\pi}{2}\right] \hfill\mbox{ for
}r<\frac{L\sin\gamma}{2},\\\\
\varphi\in\left[0,\alpha(r)\right],\left[\alpha(r),\beta(r)\right],\left[\beta(r),\frac{\pi}{2}\right]\hfill\\\\
\hfill\mbox{ for
}\frac{L\sin\gamma}{2}<r<L\sin\frac{\gamma}{2},\\\\
\varphi\in\left[0,\alpha(r)\right],\left[\alpha(r),\frac{\pi}{2}\right]
\qquad\mbox{ for
}L\sin\frac{\gamma}{2}<r<L\cos\frac{\gamma}{2},\\\\
\varphi\in\left[0,\frac{\pi}{2}\right] \hfill\mbox{ for
}r>L\cos\frac{\gamma}{2},
\end{array}
\end{equation*}
where
\begin{align}
\alpha(r) ={}& {\rm arcsin}\left(\frac{L\sin\gamma}{2r}\right) -
\frac{\gamma}{2},\\
\beta(r) ={}& \pi - {\rm
arcsin}\left(\frac{L\sin\gamma}{2r}\right) -
\frac{\gamma}{2}\label{eq:beta},
\end{align}
are the angles for which the circle intersects the boundary of the
parallelogram. In each domain, we calculate the integral
$\mathcal{A}_{l,m}$ using the corresponding expression for the
integrant $\mathcal{B}_{l}$: \eqref{eq:local} if the circle
segment lies in the interior of the parallelogram;
\eqref{eq:multi} if it lies outside of the parallelogram.

\section{The limit for parallel rods}
In principle, calculations of the effective excluded volume for
parallel rods involves the limit $\gamma\rightarrow 0$ of
Eqs.~\eqref{eq:inta}--\eqref{eq:beta}. To obtain the correct
result, one has to take care to perform the limit correctly in
each expression, which is not straightforward. It is much easier
to re-evaluate the expressions in this limit analytically,
starting with Eqs.~\eqref{eq:frame}--\eqref{eq:polar}. We use the
same reference frame, but a different substitution of variables
\begin{align}
\hat{\omega} ={}& (1,0,0),\\
\hat{\omega}' ={}& (1,0,0),\\
l\hat{\omega}-l'\hat{\omega}' ={}& (\pm x,0,0),
\end{align}
where $x=|l-l'|$. Now the integration is performed over relative
positions of two points on a single line. Half of the combinations
is positive ($l>l'$), the other half is negative ($l<l'$). The
integration boundaries of either set is given by
\begin{equation}
0 \leq x \leq L.
\end{equation}
The length over which each combination $l,l'$ is realized, for a
certain value of $x$, is given by $L-x$. In accordance with the
previous expressions, we define the integral $\mathcal{A}$ for
parallel rods as
\begin{equation}\label{eq:multip}
\mathcal{A}_{l,m}(r;\gamma=0) = 2k_{l}(\kappa r)\int_{0}^{L}{\rm
d}x\,(L-x)i_{l}(\kappa x)
\end{equation}
for $r>L$, and
\begin{align}
\mathcal{A}_{l,m}(r;\gamma=0) ={}& 2k_{l}(\kappa
r)\int_{0}^{r}{\rm d}x\,(L-x)i_{l}(\kappa x)\nonumber\\
&{}+ 2i_{l}(\kappa r)\int_{r}^{L}{\rm d}x\,(L-x)k_{l}(\kappa
x)\label{eq:localp}
\end{align}
for $r<L$. Note that the expressions are independent of $m$.

\section{Notations, integrals and Taylor series expansions}
In order to calculate the integral $\mathcal{A}$, we first need to
calculate the integral $\mathcal{B}$ by performing the
integration|over the radial coordinate $\rho$|in
Eqs.~\eqref{eq:multi} and \eqref{eq:local}. Introducing the
notation
\begin{align}
\mathcal{I}_{l}(z) ={}& \int_{0}^{z}{\rm
d}x\,x\,i_{l}(x),\label{eq:I}\\
\mathcal{K}_{l}(z) ={}& \int_{z}^{\infty}{\rm
d}x\,x\,k_{l}(x),\label{eq:K}
\end{align}
we can rewrite $\mathcal{B}$ as
\begin{align}
&\kappa^{2}\mathcal{B}_{l}(r;\varphi,\gamma)\nonumber\\
&{}=\left\{\begin{array}{l} k_{l}(\kappa
r)\mathcal{I}_{l}\left(\frac{\kappa
L\sin\gamma}{2\sin\left(\varphi+\frac{\gamma}{2}\right)}\right)
\hfill\mbox{for }
r>\frac{L\sin\gamma}{2\sin\left(\varphi+\frac{\gamma}{2}\right)},\\\\\\
k_{l}(\kappa r)\mathcal{I}_{l}(\kappa r) + i_{l}(\kappa
r)\mathcal{K}_{l}(\kappa r)\\\\
{}- i_{l}(\kappa r)\mathcal{K}_{l}\left(\frac{\kappa
L\sin\gamma}{2\sin\left(\varphi+\frac{\gamma}{2}\right)}\right)\quad\mbox{for
}
r<\frac{L\sin\gamma}{2\sin\left(\varphi+\frac{\gamma}{2}\right)}.\label{eq:B}
\end{array}\right.
\end{align}
Unfortunately, there is no (easy) way to write the expressions in
Eqs.~\eqref{eq:I} and \eqref{eq:K} explicitly for arbitrary $l$.
However, one can give explicit expressions (necessary for our
calculations) for $l=0,2,4$. First, the Bessel functions
\begin{align}
&i_{0}(z) = \frac{\sinh(z)}{z},\\
&i_{2}(z) = \frac{(z^{2}+3)\sinh(z) - 3z\cosh(z)}{z^{3}},\\
&i_{4}(z) =\nonumber\\
&\frac{(z^{4}+45z^{2}+105)\sinh(z) -
(10z^{3}+105z)\cosh(z)}{z^{5}},\\
&k_{0}(z) = \frac{\exp(-z)}{z},\\
&k_{2}(z) = \frac{(z^{2}+3z+3)\exp(-z)}{z^{3}},\\
&k_{4}(z) =
\frac{(z^{4}+10z^{3}+45z^{2}+105z+105)\exp(-z)}{z^{5}}.
\end{align}
Next, their integrals
\begin{align}
\mathcal{I}_{0}(z) ={}& \cosh(z)-1,\label{eq:I0}\\
\mathcal{I}_{2}(z) ={}& \frac{z\cosh(z) - 3\sinh(z)}{z} + 2,\\
\mathcal{I}_{4}(z) ={}& \frac{(z^{3}+35z)\cosh(z) -
(10z^{2}+35)\sinh(z)}{z^{3}} - \frac{8}{3},\\
\mathcal{K}_{0}(z) ={}& \exp(-z),\\
\mathcal{K}_{2}(z) ={}& \frac{(z+3)\exp(-z)}{z},\\
\mathcal{K}_{4}(z) ={}&
\frac{(z^{3}+10z^{2}+35z+35)\exp(-z)}{z^{3}}.\label{eq:K4}
\end{align}
Unfortunately, we cannot perform the subsequent integration|of the
angular coordinate $\varphi$|in Eq.~\eqref{eq:inta} analytically,
when we try to calculate $\mathcal{A}$. Therefore, we use the
series expansions (for even $l$)
\begin{align}
\mathcal{I}_{2n}(z) ={}&
2^{2n}\sum_{k=0}^{\infty}\frac{(2n+k)!z^{2n+2k+2}}{(2n+2k+2)(4n+2k+1)!k!},\label{eq:Is}\\
\mathcal{K}_{2n}(z) ={}&
(-1)^{n}\frac{(2^{n}n!)^{2}}{(2n)!}\nonumber\\
&{}-
\frac{1}{2^{2n}}\sum_{k=1}^{2n}\frac{(-1)^{k}(2k)!z^{2n-2k+1}}{(2n-2k+1)(2n-k)!k!}\nonumber\\
&{}-
\frac{1}{2^{2n}}\sum_{k=0}^{\infty}\frac{k!z^{2n+2k+1}}{(2n+2k+1)(2n+k)!(2k)!}\nonumber\\
&{}+
2^{2n}\sum_{k=0}^{\infty}\frac{(2n+k)!z^{2n+2k+2}}{(2n+2k+2)(4n+2k+1)!k!}.\label{eq:Ks}
\end{align}

Finally, we define the specific combination
\begin{equation}
\mathcal{C}_{l}(\kappa r) = k_{l}(\kappa r)\mathcal{I}_{l}(\kappa
r) + i_{l}(\kappa r)\mathcal{K}_{l}(\kappa r),
\end{equation}
which turns out to be given by a relatively simple expression (for
even $l$)
\begin{align}
\mathcal{C}_{2n}(\kappa r) ={}&
\frac{(n!)^{2}}{(2n)!}\sum_{k=0}^{n}\frac{(-1)^{k}(2n+2k)!}{(n+k)!(n-k)!(\kappa
r)^{2k+1}}\nonumber\\
&{} - (-1)^{n}\frac{(2^{n}n!)^{2}}{(2n)!}k_{2n}(\kappa
r),\label{eq:Cs}
\end{align}
such that
\begin{align}
\mathcal{C}_{0}(z) ={}& \frac{1}{z} - \frac{\exp(-z)}{z},\\
\mathcal{C}_{2}(z) ={}& \frac{z^{2}-6}{z^{3}} + 2\frac{(z^{2}+3z+3)\exp(-z)}{z^{3}},\\
\mathcal{C}_{4}(z) ={}& \frac{z^{4}-20z^{2}+280}{z^{5}}\nonumber\\
&{}-
\frac{8}{3}\frac{(z^{4}+10z^{3}+45z^{2}+105z+105)\exp(-z)}{z^{5}}.
\end{align}
Note that in each expression the first term cancels the divergence
of the second term in the limit where $z\rightarrow 0$. Hence,
this limit is given by
\begin{equation}
\mathcal{C}_{2n}(0) = \delta_{n,0}.
\end{equation}
This property is also reflected in the series expansion|useful for
calculations for small $\kappa r$|given by
\begin{align}
\mathcal{C}_{2n}(\kappa r) ={}&
(-1)^{n}\frac{(2^{n}n!)^{2}}{(2n)!}\nonumber\\
&{}\times
\frac{\sqrt{\pi}}{2}\sum_{k=0}^{\infty}\frac{1}{\Gamma\left(\frac{k+2n+3}{2}\right)\Gamma\left(\frac{k-2n+2}{2}\right)}\left(\frac{-\kappa
r}{2}\right)^{k}.
\end{align}
Note that the terms for even $k<2n$ have vanishing coefficients.

The limit of parallel rods has a different set of expressions.
Therefore, we define an additional notation
\begin{align}
\mathcal{J}_{l}(z) ={}& \int_{0}^{z}{\rm
d}x\,\frac{z-x}{z}\,i_{l}(x),\\
\mathcal{L}_{l}(z) ={}& \int_{z}^{\infty}{\rm
d}x\,\frac{z-x}{z}\,k_{l}(x).
\end{align}
In this way, we split each integral in Eq.~\eqref{eq:localp} in
two parts
\begin{align}
&\kappa^{2}\mathcal{A}_{l,m}(r;\gamma=0)\nonumber\\
&{}=\left\{\begin{array}{l} 2\kappa L\,k_{l}(\kappa
r)\mathcal{J}_{l}(\kappa L) \hfill\mbox{for }
r>L,\\\\\\
2\frac{L-r}{r}\,\mathcal{C}_{l}(\kappa r) + 2\kappa
L\,k_{l}(\kappa r)\mathcal{J}_{l}(\kappa r)\\\\
{}+ 2\kappa L\,i_{l}(\kappa r)(\mathcal{L}_{l}(\kappa r) -
\mathcal{L}_{l}(\kappa L)) \qquad\mbox{for } r<L.
\end{array}\right.\label{eq:A}
\end{align}
Evaluation of these integrals result in slightly more complicated
expressions, when compared to the expressions for $\mathcal{I}$
and $\mathcal{K}$ in Eqs.~\eqref{eq:I0}--\eqref{eq:K4}
\begin{align}
\mathcal{J}_{0}(z) ={}& {\rm shi}(z) + \frac{1}{z} -
\frac{\cosh(z)}{z},\label{eq:J0}\\
\mathcal{J}_{2}(z) ={}& -\frac{1}{2}{\rm shi}(z) - \frac{2}{z} +
\frac{z\cosh(z)+3\sinh(z)}{2z^{2}},\\
\mathcal{J}_{4}(z) ={}& \frac{3}{8}{\rm shi}(z) +
\frac{8}{3z}\nonumber\\
&{}- \frac{(3z^{3}+70y)\cosh(z)-(5z^{2}+70)\sinh(z)}{8z^{4}},
\end{align}
where
\begin{equation}
{\rm shi}(z) = \int_{0}^{z}{\rm d}x\,\frac{\sinh(x)}{x},
\end{equation}
is the hyperbolic sine integral.
\begin{align}
\mathcal{L}_{0}(z) ={}& \Gamma(0,z) -
\frac{\exp(-z)}{z},\\
\mathcal{L}_{2}(z) ={}& -\frac{1}{2}\Gamma(0,z) +
\frac{(z-3)\exp(-z)}{2z^{2}},\\
\mathcal{L}_{4}(z) ={}& \frac{3}{8}\Gamma(0,z) -
\frac{(3z^{3}+5z^{2}+70y+70)\exp(-z)}{8z^{4}}.\label{eq:L4}
\end{align}
In principle, one now has the exact solutions for $\mathcal{A}$ up
to $l=4$. However, we need the expressions in
Eqs.~\eqref{eq:J0}--\eqref{eq:L4} to provide a well defined limit
for the parallel rods, to use in combination with the expressions
for arbitrary orientations (i.e. the series expansions in
Eqs.~\eqref{eq:Is}, \eqref{eq:Ks}, and \eqref{eq:Cs}). Therefore,
it will be convenient to also have these expressions in the form
of a series expansion
\begin{align}
&\mathcal{J}_{2n}(z)\nonumber\\
&{}=
2^{2n}\sum_{k=0}^{\infty}\frac{(2n+k)!z^{2n+2k+1}}{(2n+2k+1)(2n+2k+2)(4n+2k+1)!k!},\\
&\mathcal{L}_{2n}(z) =
(-1)^{n}\frac{(2n)!}{(2^{n}n!)^{2}}\left(1+\sum_{k=1}^{2n}\frac{1}{k}-\gamma_{\rm E}-\ln(z)\right)\nonumber\\
&{}- (-1)^{n}\frac{(2^{n}n!)^{2}}{(2n)!}\frac{1}{z}\nonumber\\
&{}-
\frac{1}{2^{2n}}\sum_{k=0,k\neq n}^{2n}\frac{(-1)^{k}(2k)!z^{2n-2k}}{(2n-2k)(2n-2k+1)(2n-k)!k!}\nonumber\\
&{}-
\frac{1}{2^{2n}}\sum_{k=1}^{\infty}\frac{k!z^{2n+2k}}{(2n+2k)(2n+2k+1)(2n+k)!(2k)!}\nonumber\\
&{}+
2^{2n}\sum_{k=0}^{\infty}\frac{(2n+k)!z^{2n+2k+1}}{(2n+2k+1)(2n+2k+2)(4n+2k+1)!k!}.
\end{align}

\section{Truncation and some examples of expressions}
In principle, the calculation of each of the terms in
Eq.~\eqref{eq:elecsph} (i.e.~each order of $l$ and $m$) involves
an infinite series expansion in $\kappa L$. We will restrict our
calculations to $l=0,2$, and $4$, and truncate each series
expansion. Since the integration domain of $\mathcal{A}$ is shaped
like a parallelogram with sides of length $L$, we divide out a
factor $L^{2}$ to make both $\mathcal{A}$ and $\mathcal{B}$
dimensionless (i.e.~we calculate $\kappa^{2}\mathcal{A}/(\kappa
L)^{2}$ and $\kappa^{2}\mathcal{B}/(\kappa L)^{2}$). This factor
$L^{2}$ is combined with the prefactor $\kappa l_{\rm
B}\lambda^{2}$ in Eq.~\eqref{eq:elecsph}. From the definition of
the charge parameter $q$, we can write the result as an overall
prefactor $q\kappa^{2}L^{2}$. The truncated expansion is defined
as the expansion up to fourth order in $\kappa L$ of the
expression where this prefactor is taken out. This means that we
determine the series expansions of the expressions in
Eqs.~\eqref{eq:B} and \eqref{eq:A}, after we divide by a factor
$(\kappa L)^{2}$. We give some examples of the calculated
expressions for $l=0$ and $m=0$. We make the distinction between
four domains in $r$. For $r<\frac{L\sin\gamma}{2}$
\begin{align}
&\kappa^{2}\mathcal{A}_{0,0}(r;\gamma)\nonumber\\
&{}= \frac{4}{\sin\gamma}\int_{0}^{\frac{\pi}{2}}{\rm
d}\varphi\left(\mathcal{C}_{0}(\kappa r) - i_{0}(\kappa
r)\mathcal{K}_{0}\left(\frac{\kappa
L\sin\gamma}{2\sin\left(\varphi+\frac{\gamma}{2}\right)}\right)\right)\nonumber\\
&{}\simeq \frac{2\pi}{\sin\gamma}\left(\frac{1}{\kappa r} -
\frac{\exp(-\kappa r)}{\kappa r}\right) -
\frac{2\pi}{\sin\gamma}\frac{\sinh(\kappa r)}{\kappa
r}\nonumber\\
&{}+ \frac{\sinh(\kappa r)}{\kappa
r}\,\kappa^{2}L^{2}\left[-\ln\left(\tan\frac{\gamma}{4}\tan\frac{\pi-\gamma}{4}\right)\right.\nonumber\\
&\qquad\qquad\qquad\qquad{}\times \left(\frac{2}{\kappa L} +
\frac{\kappa L\sin^{2}\gamma}{24} +
\frac{\kappa^{3}L^{3}\sin^{4}\gamma}{2560}\right)\nonumber\\
&{}+ \sqrt{1+\sin\gamma}(2 - \sin\gamma)\frac{\kappa L}{12}\nonumber\\
&{}+ \sqrt{1+\sin\gamma}\left(16 - 8\sin\gamma + 2\sin^{2}\gamma -
3\sin^{3}\gamma\right)\frac{\kappa^{3}L^{3}}{3840}\nonumber\\
&\left.{}- 1 - \frac{\kappa^{2}L^{2}}{36} -
\left(7+5\cos^{2}\gamma\right)\frac{\kappa^{4}L^{4}}{21600}\right].
\end{align}

\clearpage

The next domain is
$\frac{L\sin\gamma}{2}<r<L\sin\frac{\gamma}{2}$, where the
expression gets a lot more involved
\begin{align}
&\kappa^{2}\mathcal{A}_{0,0}(r;\gamma) =
\frac{4}{\sin\gamma}\nonumber\\
&{}\times\left[\int_{0}^{\alpha(\kappa r)}{\rm
d}\varphi\left(\mathcal{C}_{0}(\kappa r) - i_{0}(\kappa
r)\mathcal{K}_{0}\left(\frac{\kappa
L\sin\gamma}{2\sin\left(\varphi+\frac{\gamma}{2}\right)}\right)\right)\right.\nonumber\\
&\quad{}+ \int_{\alpha(\kappa r)}^{\beta(\kappa r)}{\rm
d}\varphi\,k_{0}(\kappa r)\mathcal{I}_{0}\left(\frac{\kappa
L\sin\gamma}{2\sin\left(\varphi+\frac{\gamma}{2}\right)}\right)\nonumber\\
&\quad\left.{}+ \int_{\beta(\kappa r)}^{\frac{\pi}{2}}{\rm
d}\varphi\left(\mathcal{C}_{0}(\kappa r) - i_{0}(\kappa
r)\mathcal{K}_{0}\left(\frac{\kappa
L\sin\gamma}{2\sin\left(\varphi+\frac{\gamma}{2}\right)}\right)\right)\right]\nonumber\\
&{}\simeq \frac{4}{\sin\gamma}\left(2\,{\rm arcsin}(\xi) -
\frac{\pi}{2}\right)\left(\frac{1}{\kappa r} - \frac{\exp(-\kappa
r)}{\kappa r}\right)\nonumber\\
&\qquad\qquad\qquad{}- \frac{4}{\sin\gamma}\left(2\,{\rm
arcsin}(\xi) -
\frac{\pi}{2}\right)\frac{\sinh(\kappa r)}{\kappa r}\nonumber\\
&{}+ \frac{\sinh(\kappa r)}{\kappa
r}\,\kappa^{2}L^{2}\left[-\ln\left(\tan\frac{\gamma}{4}\tan\frac{\pi-\gamma}{4}\right)\right.\nonumber\\
&\qquad\qquad\qquad\qquad{}\times \left(\frac{2}{\kappa L} +
\frac{\kappa L\sin^{2}\gamma}{24} +
\frac{\kappa^{3}L^{3}\sin^{4}\gamma}{2560}\right)\nonumber\\
&{}+ \sqrt{1+\sin\gamma}(2 - \sin\gamma)\frac{\kappa L}{12}\nonumber\\
&{}+ \sqrt{1+\sin\gamma}\left(16 - 8\sin\gamma + 2\sin^{2}\gamma -
3\sin^{3}\gamma\right)\frac{\kappa^{3}L^{3}}{3840}\nonumber\\
&{}- 1 - \frac{\kappa^{2}L^{2}}{36} -
\left(7+5\cos^{2}\gamma\right)\frac{\kappa^{4}L^{4}}{21600}\nonumber\\
&{}- \frac{2r}{L}\,{\rm
arctanh}\left(\sqrt{1-\xi^{2}}\right)\left(\frac{2}{\kappa r} +
\xi^{2}\frac{\kappa r}{6} +
\xi^{4}\frac{\kappa^{3}r^{3}}{160}\right)\nonumber\\
&{}- \frac{2r}{L}\sqrt{1-\xi^{2}}\left(\frac{\kappa r}{6} +
(3\xi^{2}+2)\frac{\kappa^{3}r^{3}}{480}\right.\nonumber\\
&\qquad\left.\left.{}- 1 - (2\xi^{2}+1)\frac{\kappa^{2}r^{2}}{36}
-
(8\xi^{4}+4\xi^{2}+3)\frac{\kappa^{4}r^{4}}{5400}\right)\right]\nonumber\\
&{}+ \frac{\exp(-\kappa r)}{\kappa
r}\kappa^{2}L^{2}\left[\frac{2r}{L}\sqrt{1-\xi^{2}}\right.\nonumber\\
&{}\left.\times \left(1 + (2\xi^{2}+1)\frac{\kappa^{2}r^{2}}{36} +
(8\xi^{4}+4\xi^{2}+3)\frac{\kappa^{4}r^{4}}{5400}\right)\right].
\end{align}
We have abbreviated
\begin{equation}
\xi = \frac{L\sin\gamma}{2r}.
\end{equation}
This domain corresponds to the case where the circle of radius $r$
intersects the edge of the parallelogram twice at each quadrant.
The following domain corresponds to the case where there is just
one intersection per quadrant. Recall that we assume
$0<\gamma<\pi/2$, such that this domain is given by
$L\sin\frac{\gamma}{2}<r<L\cos\frac{\gamma}{2}$
\begin{align}
&\kappa^{2}\mathcal{A}_{0,0}(r;\gamma) =
\frac{4}{\sin\gamma}\nonumber\\
&{}\times\left[\int_{0}^{\alpha(\kappa r)}{\rm
d}\varphi\left(\mathcal{C}_{0}(\kappa r) - i_{0}(\kappa
r)\mathcal{K}_{0}\left(\frac{\kappa
L\sin\gamma}{2\sin\left(\varphi+\frac{\gamma}{2}\right)}\right)\right)\right.\nonumber\\
&\qquad\left.{}+ \int_{\alpha(\kappa r)}^{\frac{\pi}{2}}{\rm
d}\varphi\,k_{0}(\kappa r)\mathcal{I}_{0}\left(\frac{\kappa
L\sin\gamma}{2\sin\left(\varphi+\frac{\gamma}{2}\right)}\right)\right]\nonumber\\
&{}\simeq \frac{4}{\sin\gamma}\left({\rm arcsin}(\xi) -
\frac{\gamma}{2}\right)\left(\frac{1}{\kappa r} -
\frac{\exp(-\kappa
r)}{\kappa r} - \frac{\sinh(\kappa r)}{\kappa r}\right)\nonumber\\
&{}+ \frac{\sinh(\kappa r)}{\kappa
r}\,\kappa^{2}L^{2}\nonumber\\
&{}\times
\left[-\ln\left(\tan\frac{\gamma}{4}\right)\left(\frac{2}{\kappa
L} + \frac{\kappa L\sin^{2}\gamma}{24} +
\frac{\kappa^{3}L^{3}\sin^{4}\gamma}{2560}\right)\right.\nonumber\\
&{}+ \left(\frac{1+\cos\gamma}{2}\right)^{\frac{3}{2}}\frac{\kappa
L}{6} + \left(\frac{1+\cos\gamma}{2}\right)^{\frac{5}{2}}(7 -
3\cos\gamma)\frac{\kappa^{3}L^{3}}{960}\nonumber\\
&{}- \frac{1+\cos\gamma}{2} -
\left(\frac{1+\cos\gamma}{2}\right)^{2}(2-\cos\gamma)\frac{\kappa^{2}L^{2}}{36}\nonumber\\
&{}-
\left(\frac{1+\cos\gamma}{2}\right)^{3}\left(7 - 6\cos\gamma + 2\cos^{2}\gamma\right)\frac{\kappa^{4}L^{4}}{5400}\nonumber\\
&{}- \frac{r}{L}\,{\rm
arctanh}\left(\sqrt{1-\xi^{2}}\right)\left(\frac{2}{\kappa r} +
\xi^{2}\frac{\kappa r}{6} +
\xi^{4}\frac{\kappa^{3}r^{3}}{160}\right)\nonumber\\
&{}- \frac{r}{L}\sqrt{1-\xi^{2}}\left(\frac{\kappa r}{6} +
(3\xi^{2}+2)\frac{\kappa^{3}r^{3}}{480}\right.\nonumber\\
&\qquad\left.\left.{}- 1 - (2\xi^{2}+1)\frac{\kappa^{2}r^{2}}{36}
-
(8\xi^{4}+4\xi^{2}+3)\frac{\kappa^{4}r^{4}}{5400}\right)\right]\nonumber\\
&{}+ \frac{\exp(-\kappa r)}{\kappa
r}\,\kappa^{2}L^{2}\nonumber\\
&{}\times \left[\frac{1-\cos\gamma}{2} +
\left(\frac{1-\cos\gamma}{2}\right)^{2}(2+\cos\gamma)\frac{\kappa^{2}L^{2}}{36}\right.\nonumber\\
&{}+ \left(\frac{1-\cos\gamma}{2}\right)^{3}\left(7 + 6\cos\gamma
+ 2\cos^{2}\gamma\right)\frac{\kappa^{4}L^{4}}{5400}\nonumber\\
&{}+ \frac{r}{L}\sqrt{1-\xi^{2}}\left(1 + (2\xi^{2}+1)\frac{\kappa^{2}r^{2}}{36}\right.\\
&\qquad\qquad\qquad\qquad\qquad\qquad\left.\left.{}+
(8\xi^{4}+4\xi^{2}+3)\frac{\kappa^{4}r^{4}}{5400}\right)\right].\nonumber
\end{align}
Finally, the domain where $r>L\cos\frac{\gamma}{2}$ yields a more
friendly expression
\begin{align}
&\kappa^{2}\mathcal{A}_{0,0}(r;\gamma)\nonumber\\
&{}= \frac{4}{\sin\gamma}\int_{0}^{\frac{\pi}{2}}{\rm
d}\varphi\,k_{0}(\kappa r)\mathcal{I}_{0}\left(\frac{\kappa
L\sin\gamma}{2\sin\left(\varphi+\frac{\gamma}{2}\right)}\right)\nonumber\\
&{}\simeq \frac{\exp(-\kappa r)}{\kappa r}\,\kappa^{2}L^{2}\left[1
+ \frac{\kappa^{2}L^{2}}{36} +
\left(7+5\cos^{2}\gamma\right)\frac{\kappa^{4}L^{4}}{21600}\right].
\end{align}

In the case of parallel rods, we can apply the alternative series
expansions, or apply the limit $\gamma\rightarrow 0$ on the last
two expressions above. Both yield the following approximations,
where for $r<L$
\begin{align}
&\kappa^{2}\mathcal{A}_{0,0}(r;\gamma=0)\nonumber\\
&{}= 2\,\frac{L-r}{r}\left(\frac{1}{\kappa r} - \frac{\exp(-\kappa
r)}{\kappa r}\right)\nonumber\\
&{}+ 2\kappa L\,\frac{\exp(-\kappa r)}{\kappa r}\left({\rm
shi}(\kappa r) + \frac{1}{\kappa r} - \frac{\cosh(\kappa
r)}{\kappa r}\right)\nonumber\\
&{}+ 2\kappa L\,\frac{\sinh(\kappa r)}{\kappa
r}\left(\Gamma(0,\kappa
r) - \frac{\exp(-\kappa r)}{\kappa r}\right.\nonumber\\
&\qquad\qquad\qquad\qquad\left.{}- \Gamma(0,\kappa L) +
\frac{\exp(-\kappa L)}{\kappa L}\right)\nonumber\\
&{}\simeq 2\,\frac{L-r}{r}\left(\frac{1}{\kappa r} -
\frac{\exp(-\kappa
r)}{\kappa r}\right)\nonumber\\
&{}+ 2\kappa L\,\frac{\exp(-\kappa r)}{\kappa r}\frac{\kappa
r}{2}\left(1 + \frac{\kappa^{2}r^{2}}{36} +
\frac{\kappa^{4}r^{4}}{1800}\right)\nonumber\\
&{}+ 2\kappa L\,\frac{\sinh(\kappa r)}{\kappa
r}\left[\ln\left(\frac{L}{r}\right) - \frac{1}{\kappa
L}\frac{L-r}{r}\right.\nonumber\\
&\qquad{}+ \frac{\kappa r}{2}\left(1 - \frac{\kappa r}{6} +
\frac{\kappa^{2}r^{2}}{36} - \frac{\kappa^{3}r^{3}}{240} +
\frac{\kappa^{4}r^{4}}{1800}\right)\nonumber\\
&\qquad\left.{}- \frac{\kappa L}{2}\left(1 - \frac{\kappa L}{6} +
\frac{\kappa^{2}L^{2}}{36} - \frac{\kappa^{3}L^{3}}{240} +
\frac{\kappa^{4}L^{4}}{1800}\right)\right].
\end{align}
For $r>L$ we obtain
\begin{align}
&\kappa^{2}\mathcal{A}_{0,0}(r;\gamma=0)\nonumber\\
&{}= 2\kappa L\,\frac{\exp(-\kappa r)}{\kappa r}\left({\rm
shi}(\kappa L) + \frac{1}{\kappa L} - \frac{\cosh(\kappa
L)}{\kappa L}\right)\nonumber\\
&{}\simeq 2\kappa L\,\frac{\exp(-\kappa r)}{\kappa r}\frac{\kappa
L}{2}\left(1 + \frac{\kappa^{2}L^{2}}{36} +
\frac{\kappa^{4}L^{4}}{1800}\right).
\end{align}

Likewise, there are expressions for $l=2,4$. These are all used
together to create an (approximate) expression for the pair
interaction outside of the hard-core exclusion region. We use this
pair interaction to numerically calculate the effective excluded
volume. This is accomplished by a numerical integration scheme
over all different domains of $r$, for given rod orientations. Our
approach is fundamentally different from other theoretical work
\citep{Ramirez,Trizac}, in the sense that we apply the interchange
of the two positional vectors $\mathbf{r}$ and
$l\hat{\omega}-l'\hat{\omega}'$. We have to do this in order to
calculate the full integral over $r$, in contrast to the studies
in Refs.~\citep{Ramirez,Trizac}, where only a description is given
of the pair interaction for rods at large distances. Conversely,
if one considers non-spherical charge distributions on spherical
particles, this switch is not needed when introducing rotational
invariants.

\end{document}